\documentclass[twocolumn,amsfonts,amsmath,amssymb,aps,prc,10pt,floatfix,showpacs,superscriptaddress]{revtex4-1}
\usepackage[english]{babel}
\usepackage{bm}
\usepackage{natbib}
\usepackage{hyphenat}
\usepackage[utf8]{inputenc}	%Hvis du benytter windows i stedet for linux, så skift utf8 ud med latin1. Tillader danske tegn.
\usepackage[T1]{fontenc}	%Tillader danske tegn
\usepackage{graphicx}		%Tillader indsættelse af billeder
\usepackage{booktabs}		%linjer i tabeller...
\usepackage{mathtools}		%Ekstra matematik... bare lad den være, du får muligvis brug for den.
\usepackage{siunitx}	%Bruges \SI{<tal>}{<enhed>}, \si{<enhed>} eller \num{<tal>}.
\usepackage{url}		 %bruges til at formattere url'er... kan sagtens udelades.
\newcommand{\tref}[1]{\tablename~\ref{#1}}
\newcommand{\fref}[1]{\figurename~\ref{#1}}
\usepackage{hyperref}
\hypersetup
{   pdfsubject={Nuclear Physics},
	pdfauthor={Jonas Refsgaard}
    pdftitle={Spectroscopy on 11B},
    pdfstartview=,%FitH,
    colorlinks=true}
    
\makeatletter
\begingroup
\catcode`\_=\active
\protected\gdef_{\@ifnextchar|\subtextup\sb}
\endgroup
\def\subtextup|#1|{\sb{\textup{#1}}}
\AtBeginDocument{\catcode`\_=12 \mathcode`\_=32768 }
\makeatother

\usepackage[]{csquotes} %Danske citationstegn. \enquote{}
\usepackage{bm}

\begin{document}

\title{Clarification of large-strength transitions in the \texorpdfstring{$\bm\beta$}{beta} decay of \texorpdfstring{$^{\mathbf{11}}\mathbf{Be}$}{11Be}}
\author{J.~Refsgaard}\email{jonas.refsgaard@kuleuven.be}%Forfatter 1
\affiliation{Instituut voor Kern- en Stralingsfysica, KU Leuven, B-3000 Leuven, Belgium}
\author{J.~B\"uscher}
\affiliation{Instituut voor Kern- en Stralingsfysica, KU Leuven, B-3000 Leuven, Belgium}
\author{A.~Arokiaraj}
\affiliation{Instituut voor Kern- en Stralingsfysica, KU Leuven, B-3000 Leuven, Belgium}
\author{H.O.U.~Fynbo}
\affiliation{Institut for Fysik og Astronomi, Aarhus Universitet, DK-8000 Aarhus, Denmark}
\author{R.~Raabe}
\affiliation{Instituut voor Kern- en Stralingsfysica, KU Leuven, B-3000 Leuven, Belgium}
\author{K.~Riisager}
\affiliation{Institut for Fysik og Astronomi, Aarhus Universitet, DK-8000 Aarhus, Denmark}
\date{\today} %Dato. Husk at ændre!

\begin{abstract}
The shape and normalisation of the $\beta$-delayed $\alpha$ spectrum from ${^{11}\mathrm{Be}}$ was measured by implanting ${^{11}\mathrm{Be}}$ ions in a segmented Si detector. The spectrum is found to be dominated by a well-known transition to the $3/2^+$ state at $E_x = \SI{9.87}{\mega\electronvolt}$ in ${^{11}\mathrm{B}}$. A significant increase in the observed decay strength towards the higher end of the $Q_\beta$ window means, however, that the \SI{9.87}{\mega\electronvolt} state cannot alone be responsible for the transition. Using the $R$-matrix framework we find that the inclusion of an extra $3/2^+$ state at $E_x=\SI{11.49 \pm 0.10}{\mega\electronvolt}$ is required in order to obtain a satisfactory description of the spectrum. Both states show large widths towards $\alpha$ decay, exhausting significant fractions of the Wigner limit, a typical signature of $\alpha$ clusterisation. The observed Gamow-Teller strength indicate large overlaps between the two states and the ground state of ${^{11}\mathrm{Be}}$.
\end{abstract}

\pacs{21.10.Gv, 21.10.Re, 21,10.Tg, 23.40.-s, 27.20.+n} %OBS: Change!

\maketitle

\section{Introduction}
\begin{figure}[htbp]
\centering
\includegraphics[width=0.85\columnwidth]{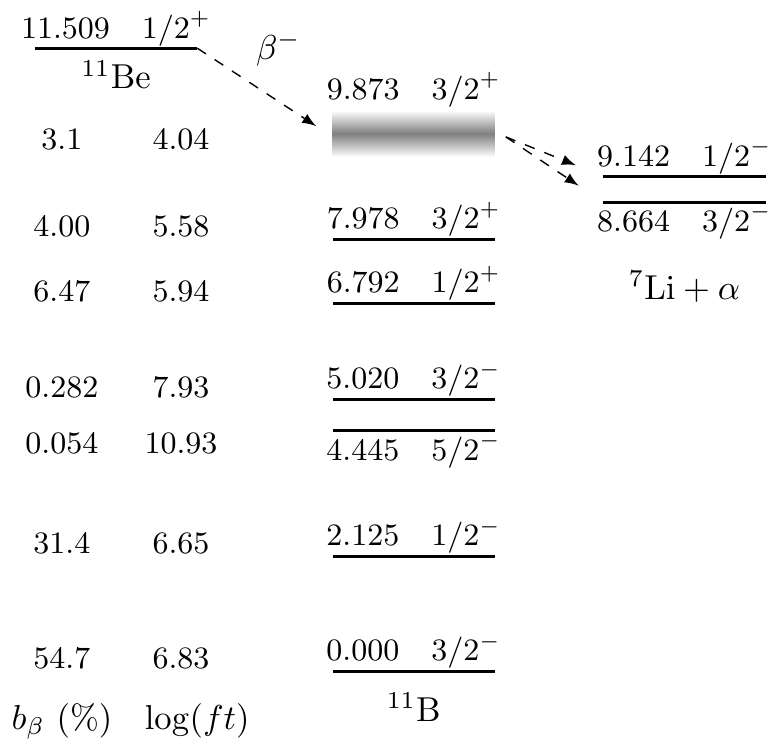}
\caption{Decay scheme of ${^{11}\mathrm{Be}}$. Each level is labelled by its energy in \si{\mega\electronvolt} relative to the ${^{11}\mathrm{B}}$ ground state as well as its spin and parity, $J^\pi$. Also included are the known $\beta$-decay branching ratios and corresponding $\log(ft)$ values. The values are based on \cite{Kelley2012} and the work presented in Refs.~\cite{Alburger1981,Millener1982}.}
\label{fig:scheme}
\end{figure}

Halo nuclei form a class of clusterised nuclear systems characterised by having one or two nucleons in very extended orbitals around the remaining nucleons. The ${^{11}\mathrm{Be}}$ ground state is a typical example of a halo system and is well described as a ${^{10}\mathrm{Be}}$ core coupled to an $s$-wave halo neutron~\cite{Riisager2013}. The halo structure influences the fundamental properties of this state, which furthermore has the unusual spin and parity $J^\pi = 1/2^+$~\cite{Kelley2012}, contrary to the $1/2^-$ designation which would be expected from the shell model. This has an important impact on the $\beta$ decay from ${^{11}\mathrm{Be}}$ to ${^{11}\mathrm{B}}$ (see \fref{fig:scheme}), where the lowest four levels have negative parity. The transitions to these levels are therefore all first-forbidden and strongly suppressed. Also, the allowed transitions to the \SI{6.79}{\mega\electronvolt}- and \SI{7.98}{\mega\electronvolt}-states are quite slow, with $\log(ft)$ values of \num{5.94} and \num{5.58}, respectively. The hindrance of these transitions not only results in ${^{11}\mathrm{Be}}$ having an unusually long half-life (\SI{13.81}{\second}~\cite{Alburger1970}), but also leads to a relative enhancement of branches with a small $Q_\beta$. This has recently allowed the detection of the rather exotic $\beta^- p$ decay branch~\cite{Riisager2014b} and it makes ${^{11}\mathrm{Be}}$ ideal for studying the hypothesised dark decay of the neutron~\cite{Fornal2018,Pfutzner2018}.

%On the other hand, the hindrance of those transitions with the largest $Q_\beta$ values is equivalent to a relative enhancement of transitions to the upper part of the $Q_\beta$ window. Since a major part of the Gamow-Teller (GT) strength is expected to appear around the same energy as the mother state~[??], this enhancement means that the ${^{11}\mathrm{Be}}$ decay is ideal for studying $\beta$ transitions with large GT matrix elements. If the states in the daughter nucleus can be sufficiently well characterised the distribution of the GT strength might even give hints about the structure of the mother state.
%

The upper part of the $Q_\beta$ window lies above the threshold for $\alpha$-particle emission ($E_|thres| = \SI{8.664}{\kilo\electronvolt}$), and can therefore be studied by observing the spectrum of $\beta$-delayed $\alpha$ particles. The spectrum was most recently measured by Alburger \textit{et al.}~\cite{Alburger1981} several decades ago. In that study it was concluded that the allowed transition to the $3/2^+$ state at \SI{9.87}{\mega\electronvolt} was alone responsible for the delayed alphas. Alburger \textit{et al.} also made a couple of observations regarding the shape of the \SI{9.87}{\mega\electronvolt} peak, namely that the width observed in $\beta$ decay is significantly larger than that observed in scattering and reaction experiments~\cite{Li1954,Bichsel1957,Hinds1959,Cusson1966,Zwieglinski1982}, and that there are signs of an unexplained increase in the spectrum near the very top of the $Q_\beta$ window (see Figure 5 of Ref.~\cite{Alburger1981}). Due to ill-determined experimental resolution and poor statistics, the authors did not attach any significance to these features. It is one goal of the present work to address the open questions left by the earlier experiment.

We describe a measurement of the shape and normalisation of the $\beta$-delayed $\alpha$ spectrum from ${^{11}\mathrm{Be}}$. The experimental technique involves implanting ${^{11}\mathrm{Be}}$ ions in a finely segmented Si-detector and counting the implantations and decays. This method has successfully been used in several other cases to measure $\beta$-delayed particles, even very weak branches, from $^6\mathrm{He}$~\cite{Smirnov2005,Raabe2009}, ${^{11}\mathrm{Li}}$\cite{Raabe2008}, ${^{12}\mathrm{B}}$ and ${^{12}\mathrm{N}}$~\cite{Hyldegaard2009}, ${^8\mathrm{B}}$~\cite{Roger2012} and ${^{16}\mathrm{N}}$~\cite{Refsgaard2016}. The $\beta$ transition and the populated daughter states in ${^{11}\mathrm{B}}$ are characterised by fitting an $R$-matrix expression to the observed spectrum and deriving values for the reduced Gamow-Teller (GT) matrix elements, as well as level energies and partial widths. Lastly, we obtain accurate and precise values of the branching ratio for delayed $\alpha$ decay, $b_{\beta\alpha}$, and the intensity of the \SI{478}{\kilo\electronvolt} $\gamma$ line emitted from the excited ${^7\mathrm{Li}}$ fragment. These values will provide two new methods for reliable normalisation of the ${^{11}\mathrm{Be}}$ decay, one relying solely on detection of charged particles.
%This value will allow for reliable normalisation of future ${^{11}\mathrm{Be}}$ decay experiments from detection of charged particles alone.

\section{\label{sec:experiment}Experiment}
The measurement was performed by implanting a ${^{11}\mathrm{Be}}$ beam in a finely segmented double-sided silicon strip detector (DSSSD) and counting the implantations and decays. The ${^{11}\mathrm{Be}}$ beam was produced at CERN's ISOLDE facility~\cite{Borge2018} by bombarding a Ta target with \SI{1.3}{\micro\ampere} of \SI{1.4}{\giga\electronvolt} protons. The cocktail of isotopes resulting from this bombardment was in stages surface ionised in a tungsten tube and in the resonant ionisation laser ion source (RILIS~\cite{Mishin1993}) and subsequently separated in the general purpose mass separator. An energetic beam of ${^{11}\mathrm{Be}^{4+}}$ ions was produced in the REX-ISOLDE post-acceleration stage~\cite{Kester2003}: The ions were injected and stored in a Penning trap (REXTRAP) from which bunches of ions were transported to an electron beam ion source (EBIS) for charge breeding. With a frequency of $\sim\SI{39}{\hertz}$ the EBIS injected ion bunches into the REX linear accelerator which accelerated the ions to an energy of \SI{2.86}{\mega\electronvolt\per A}, corresponding to a total kinetic energy of \SI{31.46}{\mega\electronvolt} per ion. At this point the beam still contained some contamination from stable ${^{22}\mathrm{Ne}}$ ions. This contamination was removed by placing Al foils with a combined thickness of \SI{36}{\micro\meter} in front of our detection system. These foils completely stopped the ${^{22}\mathrm{Ne}}$ ions while still leaving the ${^{11}\mathrm{Be}}$ ions with $\sim\SI{15}{\mega\electronvolt}$ of kinetic energy.

The radioactive beam was implanted in a DSSSD with a thickness of \SI{78}{\micro\meter} and an active area of \SI[product-units = power, input-quotient = :, output-quotient = \text{--}, quotient-mode = symbol]{16 x 16}{\mm}~\cite{Sellin1992}. Both sides of the DSSSD are covered by 48 \SI{300}{\micro\meter} wide strips with a \SI{35}{\micro\meter} interstrip spacing, resulting in a total of 2304 pixels. One particular advantage of such a fine segmentation is that the $\beta$ particles only deposit a small amount of energy in the active volume of the pixel, while the delayed particles are typically completely stopped within a single pixel. Other aspects of the implantation technique are discussed in Ref.~\cite{Buscher2008}. The implantation depth of the radioactive ions is calculated using the LISE++ toolbox~\cite{Tarasov2016} to be \SI{25}{\micro\meter}. The highest-energy $\alpha$ particles from the $\beta$-delayed $\alpha$ decay of ${^{11}\mathrm{Be}}$ are emitted with a kinetic energy of \SI{1.8}{\mega\electronvolt}, giving them a range of \SI{6.4}{\micro\meter} in Si. We therefore assume that all the $\alpha$ particles from the decay are stopped within the detector.

The strips on each side of the DSSSD were connected to a Mesytec MPR64 pre-amplifier. The signals from the front strips were then split and fed to two sets of Mesytec STM16+ amplifiers: One set with a high gain setting for spectroscopy on the delayed particles, and one set with a lower gain for identification of implantation events. The signals from the back strips were only fed to one set of STM16+ amplifiers with a high gain setting. The amplified signals were digitised using CAEN V785 ADCs. The trigger level on the low gain amplifiers was set at $\sim\SI{8}{\mega\electronvolt}$ and the trigger signal was fed to a CAEN V830 scaler in order to count the number of implantation events. The dead time was monitored by counting both the total and accepted number of triggers in the scaler. The entire setup was operated in beam on/off mode with each spill from the EBIS triggering a \SI{1}{\milli\second} beam-on gate. The total duration of the experiment was $T_|tot| = \SI{3141.5}{\second}$

\section{Analysis}
\subsection{Event reconstruction}
Following the experiment the detector-amplifier-ADC system was energy calibrated using an external calibration source containing ${^{148}\mathrm{Gd}}$, ${^{239}\mathrm{Pu}}$, ${^{241}\mathrm{Am}}$ and ${^{244}\mathrm{Cm}}$. The $\alpha$-particle energies from Ref.~\cite{Rytz1991} are corrected for energy loss through the detector dead layer, which has been measured to be \SI{340 \pm 3}{\nano\meter} thick~\cite{Buscher2008}. The ADC channel corresponding to zero signal is also registered and used as a point in the energy calibration.

\begin{figure}[htbp]
\centering
\includegraphics[width=\columnwidth]{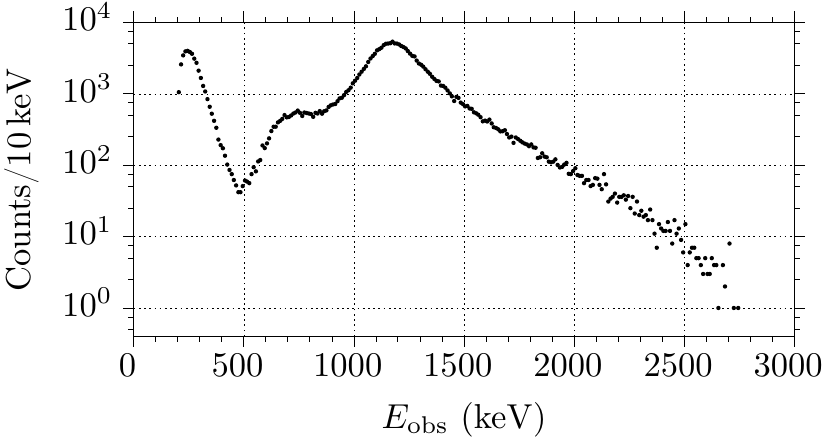}
\caption{Observed decay spectrum resulting from the reconstruction procedure described in the text.}
\label{fig:data}
\end{figure}

The decay of an implanted ${^{11}\mathrm{Be}}$ ion produces a signal in both the front and the back side of the DSSSD. Ideally this will result in an event where the signal is carried away by a single front strip and a single back strip. In this type of event, the decay energy is taken as the average of the front and back signals. If, however, the energy is deposited between two strips, i.e. in one of the interstrip regions, the signal is divided between the two neighbouring strips. This phenomenon is known as charge sharing, and from simple, geometric considerations we expect that $\sim\SI{20}{\percent}$ of the decays will suffer from the effect. To recover events with charge sharing in one or both sides of the DSSSD we also include events with signals in two neighbouring strips in the analysis by using their combined signal. We find that this correction increases the number of observed decays by $\sim\SI{17}{\percent}$, in reasonable agreement with the geometric estimate.
%Since $\beta$ particles can travel large distances in silicon they also have a tendency to produce signals in several adjacent strips, and our correction for charge sharing therefore increases the contribution from pure $\beta$ decays in the spectrum. To reduce the pure $\beta$ signal, we introduce an energy threshold and demand that a charge sharing event must have a signal above this threshold in at least one of the strips. A threshold value of \SI{350}{\kilo\electronvolt} seemed to provide a good suppression of the $\beta$ signal whithout affecting the particle spectrum.
%Finally, to ensure that the front and back signals originate from the same decay, we also demand $\lvert E_|front| - E_|back| \rvert < \SI{160}{\kilo\electronvolt}$.
Finally, to ensure that the front and back signals originate from the same decay, we match those signals that give the minimum $\lvert E_|front| - E_|back| \rvert$. The resulting spectrum is shown in \fref{fig:data}.

We see the pure $\beta$ signal producing a sharp increase in the spectrum towards lower energies. At \SI{500}{\kilo\electronvolt} the particle decays start to dominate the spectrum, showing one dominant peak from ${^7\mathrm{Li(gs)}} + \alpha$ decays and a smaller peak corresponding to ${^7\mathrm{Li}(478)} + \alpha$ decays. The integral above \SI{500}{\kilo\electronvolt} is \num{1.802e5}, which we take as the observed number of particle decays. Since the particle spectrum and the $\beta$ spectrum overlap, this figure comes with a possible systematic uncertainty, which we estimate to be on the order of a few hundred counts, comparable to the statistical uncertainty. The spatial distribution of the $\alpha$ decays across the detector surface is shown in \fref{fig:patterns}(a).

\subsection{\label{sec:branching}Branching ratio}
In order to determine the absolute branching ratio for $\beta$-delayed $\alpha$ decays, $b_{\beta\alpha}$, we must, in addition to the number of $\alpha$ decays, determine the number of implanted ${^{11}\mathrm{Be}}$ ions. This is done by monitoring the trigger signals from the low-gain amplifier chain (see Sec.~\ref{sec:experiment}). In practice this means that all signals in the front strips corresponding to an energy higher than \SI{8}{\mega\electronvolt} are counted as implantations.
\begin{figure}[htbp]
\centering
\includegraphics[width=\columnwidth]{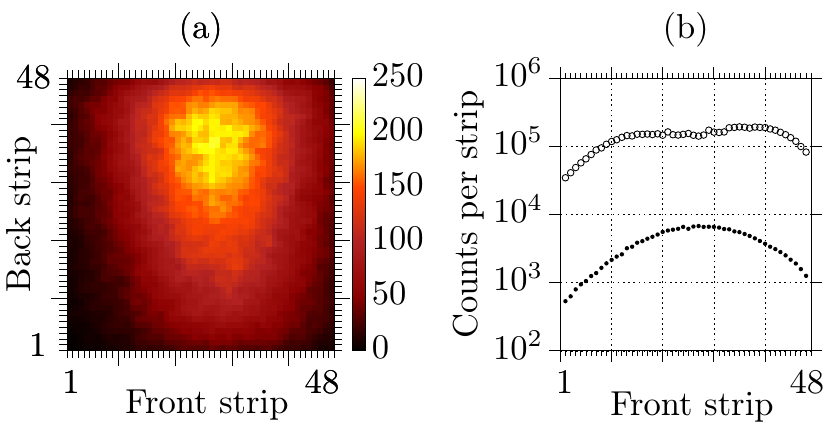}
\caption{(a) Decay pattern from events with $E_|obs|>\SI{500}{\kilo\electronvolt}$, which we identify as $\alpha$ decays. (b) Profile of the decay events (dots) and implantation events (circles) along the front strips.}
\label{fig:patterns}
\end{figure}
The implantation pattern is shown together with the decay pattern in \fref{fig:patterns}(b). In principle, the implantations and decays should follow the same spatial distribution, however, we clearly see a difference between the two patterns, with the implantation pattern showing some sort of saturation effect. We expect this to be a pile-up effect due to the very large implantation rates during the EBIS spills.

The pile-up unfortunately means that it is not possible to use all our statistics to calculate $b_{\beta\alpha}$. When we compare the two profiles in \fref{fig:patterns}(b), we find that the ratio between them is approximately constant in the first seven front strips (within the statistical uncertainty). Since the implantation rate already rises by a factor \num{2.6} from strip 1 to strip 7, we believe this shows that the implantation rate in these strips is sufficiently low that pile-up effects are not important. In order to avoid the pile-up effects as much as possible, we restrict ourselves to determine $b_{\beta\alpha}$ by using data from only front strips 1, 2 and 3.

The time structure of the data is governed by the \SI{39}{\hertz} cycle of the EBIS spills. The actual implantation events happen in a very short time interval, and the duration of the following beam-off period is approximately \SI{25}{\milli\second}. This is much shorter than the \SI{13.81}{\second} half-life of ${^{11}\mathrm{Be}}$, and the decay rate of the implanted activity is assumed to be constant during the beam-off period. To determine the dead time of the data acquisition the number of trigger requests and accepted triggers are counted for each beam-off period. To do this requires knowledge about the beginning and end of each beam-off period, which is obtained using trigger signals from the EBIS control. If these signals arrive while the data acquisition is busy, they are missed, and the decays can not be properly normalised. To validate data from a particular beam-off period we require that its first and last event happen within the duration of an EBIS cycle. This selection criterion reduces the effective measurement time to $T_|eff| = \SI{1379.5}{\second}$. From the number of trigger requests and the number of accepted triggers we derive an average detection efficiency during beam-off of $\epsilon = \num{0.883 \pm 0.001}$. The observed number of $\alpha$ decays in front strips 1-3 is $N_\alpha = 1585$ while the number of implantations in the same strips is $N_|impl| = 124100$. The resulting branching ratio for $\beta$-delayed $\alpha$ decay is
\begin{align}
b_{\beta\alpha} = \frac{N_\alpha}{\epsilon N_|impl|}\frac{T_|tot|}{T_|eff|} = \SI{3.30 \pm 0.10}{\percent}
\end{align}
where the uncertainty is estimated from the statistical uncerainty on $N_\alpha$.

\subsection{Spectral analysis}
Before it is possible to compare the observed decay spectrum with any theoretical model, we must take into account all the experimental effects, which can distort the observed signal. We consider the following three effects:
\begin{itemize}
\item $\beta$ summing.
\item Finite detector resolution.
\item The different pulse heights produced by the $\alpha$ and $^7\mathrm{Li}$ fragments from the $^{11}\mathrm{B}$ breakup.
\end{itemize}
Below we explain each of these effects and discuss how they can be corrected for.

The dominating distortion of the decay spectrum is due to the effect of $\beta$ summing, which is a result of the particle breakup being immediately preceded by the emission of of a $\beta$ particle. In almost all cases the $\beta$ particle leaves the active volume of the detector completely, however, on its way out it unavoidably creates a small signal, which is summed with the signal from the following particle breakup. The magnitude of the $\beta$ summing is dependent on the travelled distance in the detector as well as the initial kinetic energy of the $\beta$ particle.

To characterise the $\beta$ summing we simulate $10^8$ $\beta$ decays throughout the window of $Q_\beta$ values that are open to $\alpha$ emission: First, the spatial decay point is selected. We assume the decaying ions to be implanted in a depth of \SI{25}{\micro\meter} and to be uniformly distributed in the transverse directions. Second, the kinetic energy of the $\beta$ particle is chosen by sampling the allowed $\beta$ spectrum for the relevant $Q_\beta$. Third, the direction of the emitted $\beta$ particle is chosen from an isotropic distribution. Using the GEANT4 simulation toolkit~\cite{Agostinelli2003} we then track and record the energy loss of the $\beta$ particle until it has left the active detector volume. We find the most probable energy loss to be in the \SIrange{10}{15}{\kilo\electronvolt} range, however, the energy-loss distribution has a significant tail towards higher energy, and in rare events the energy loss may be several hundreds of \si{\kilo\electronvolt}.

The finite resolution of our detector also affects the observed spectrum. From the observed width of the calibration alpha peaks we estimate the resolution to be $\mathrm{FWHM} = \SI{24 \pm 5}{\kilo\electronvolt}$. This figure is dominated by the electronic resolution, which is observed to remain constant during the experiment. In the analysis we assume the detector to have a Gaussian response with an energy independent resolution of $\mathrm{FWHM} = \SI{24}{\kilo\electronvolt}$.

When the unbound state in $^{11}\mathrm{B}$ breaks apart, $\frac{7}{11}$ of the available energy is imparted on to the $\alpha$ particle and $\frac{4}{11}$ is carried away by the $^{7}\mathrm{Li}$ ion. It is well known that the efficiency for converting kinetic energy to electron-hole pairs is dependent on $Z$~\cite{Lennard1986}, and it is therefore strictly speaking a mistake to apply a calibration based on the measurement of $\alpha$ particles to the signal from $^7\mathrm{Li}$ ions. A method is given in Ref.~\cite{Kirsebom2014} to calculate the error, which is made when applying a calibration performed with particles of type $A$ to signals created by particles of type $B$. The error is written $\mathcal{E} = E_{n,A} - \Delta_{AB}$, where $E_{n,A}$ is the energy which is lost by particle A to non-ionising processes, and which is therefore not available to producing electron-hole pairs. $\Delta_{AB}$ is defined in eq.~(5) of Ref. \cite{Kirsebom2014}.
\begin{figure}[htbp]
\centering
\includegraphics[width=\columnwidth]{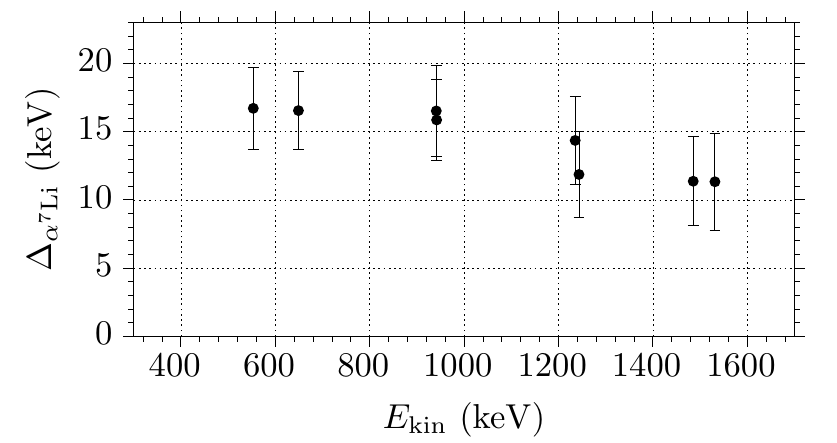}
\caption{Values of $\Delta_{\alpha {^7\mathrm{Li}}}$ calculated using the methods and results in Refs. \cite{Lennard1986,Kirsebom2014}. $E_|kin|$ refers to the kinetic energy of the $^7\mathrm{Li}$ ion when it enters the active volume of the detector.}
\label{fig:delta}
\end{figure}
The response of Si detectors to both $\alpha$ particles and $^7\mathrm{Li}$ ions were measured by Lennard \textit{et al.}~\cite{Lennard1986}, and we use the values from their Table 1 to calculate $\Delta_{\alpha {^7\mathrm{Li}}}$. The result is shown in \fref{fig:delta}. The typical kinetic energy of the $^7\mathrm{Li}$ ions is in our experiment around \SI{450}{\kilo\electronvolt}, which is somewhat outside the range of the data in \fref{fig:delta}, however, since the values seem to change only slowly with energy, we believe that it is reasonable to apply $\Delta_{\alpha {^7\mathrm{Li}}} = \SI{17 \pm 3}{\kilo\electronvolt}$ in our analysis. From Lennard \textit{et al.}~\cite{Lennard1986} we also find $E_{n,\alpha}=\SI{9 \pm 1}{\kilo\electronvolt}$, a number which is almost independent of the $\alpha$ particle energy. The expected error is then $\mathcal{E} = \SI{9 \pm 1}{\kilo\electronvolt} - \SI{17 \pm 3}{\kilo\electronvolt} = \SI{-8 \pm 3}{\kilo\electronvolt}$. From this result we conclude that $^7\mathrm{Li}$ ions create electron-hole pairs less efficiently than $\alpha$ particles, and that the net effect of applying an $\alpha$ energy calibration is that the observed decay spectrum is shifted \SI{8}{\kilo\electronvolt} towards lower energies.

Eventually, we need to calculate the observed spectrum from a theoretical model. In order to account for the effects described in the foregoing paragraphs, we construct a response matrix, $\mathbf{M}_r$, where each column contains the normalised response function for the corresponding energy bin. The observed spectrum, $\mathbf{n}_|obs|$, is then calculated from the model, $\mathbf{n}_|model|$ through a simple matrix multiplication:
\begin{align}
\mathbf{n}_|obs| = \mathbf{M}_r \mathbf{n}_|model|
\end{align}
We use the results of the GEANT4 simulation to construct a response matrix for the $\beta$ summing, $\mathbf{M}_\beta$, and a Gaussian function with $\mathrm{FWHM} = \SI{24}{\kilo\electronvolt}$ and mean $\mu = \SI{-8}{\kilo\electronvolt}$ to construct a response matrix for the detector system, $\mathbf{M}_d$. The total response matrix is obtained by $\mathbf{M}_r = \mathbf{M}_d \mathbf{M}_\beta$.
\begin{figure}[htbp]
\centering
\includegraphics[width=\columnwidth]{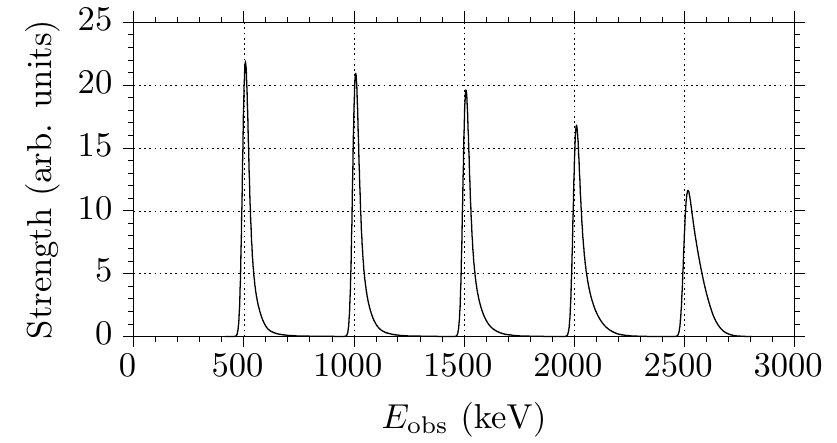}
\caption{Observed spectrum from a hypothetical model with delta-function peaks at \SI{500}{\kilo\electronvolt}, \SI{1000}{\kilo\electronvolt}, ... \SI{2500}{\kilo\electronvolt}. Energy dependent peak shifts and asymmetric broadening are visible.}
\label{fig:response}
\end{figure}
In order to visualise the net result, we have calculated the observed spectrum from a hypothetical model spectrum, consisting of five delta functions at \SI{500}{\kilo\electronvolt}, \SI{1000}{\kilo\electronvolt}, ... up to \SI{2500}{\kilo\electronvolt}, see \fref{fig:response}. We see that, even though the peaks all have the same area, the heights and widths change visibly and, in particular at high energies, the distortion is quite severe. This observation is also what we should intuitively expect, since at high $\alpha$ decay energies, $Q_\beta$, and therefore also the average kinetic energy of the $\beta$ particles, becomes very small. The stopping power of slow electrons is larger than for electrons with several \si{\mega\electronvolt} of kinetic energy, and therefore the slow $\beta$ particles tend to deposit a larger signal in the detector.

We analyse the spectrum using the phenomenological $R$-matrix method~\cite{Lane1958,Vogt1962,Descouvemont2010}. In this framework, the compound nucleus is described in terms of a number of levels, which we denote by the index $\lambda$. Each level is characterised by its energy, $E_\lambda$, and its reduced width amplitudes, $\gamma_{\lambda c}$, which determine the partial width for decay of the level $\lambda$ through channel $c$, where $c$ defines both the state of the emitted fragments, as well as their relative angular momentum. To calculate the $\beta$-delayed particle spectrum, we follow Barker and Warburton~\cite{Barker1988} and write the spectral density as
\begin{align}
\label{eq:spectrum}
N_c(E) = f_\beta P_c \Big\lvert \sum_{\lambda \mu} B_{\lambda} \gamma_{\mu c} A_{\lambda \mu} \Big\rvert^2 .
\end{align}
Here, $f_\beta$ is the phase-space factor for the $\beta$ decay, $P_c$ is the barrier penetrability and $B_\lambda$ is the $\beta$-decay feeding factor of level $\lambda$ (note that we only consider Gamow-Teller decays). The quantity $A_{\lambda \mu}$ is an element of the level matrix, which can be calculated in several ways. We follow the method presented in Ref.~\cite{Brune2002}, since it allows us to use \enquote{on-resonance} parameters directly in eq.~\eqref{eq:spectrum}. This choice makes it particularly simple to connect the formal $R$-matrix parameters to observed quantities.

The practical implementation of eq. \eqref{eq:spectrum} involves evaluation of the $\beta$-decay phase-space factor as well as the regular and irregular Coulomb wave functions. For the former we use the approximation provided by Wilkinson and Macefield~\cite{Wilkinson1974}, and for the latter we use the numerical method presented by Michel~\cite{Michel2007}. Our implementation also relies on numerical routines provided by the ROOT library~\cite{Brun1997}.

To fit eq. \eqref{eq:spectrum} to our data, we first calculate the model spectrum for particular values of the fitting parameters, modify the spectrum with the appropriate response matrix and evaluate the Poisson likelihood chi-square~\cite{Baker1984}
\begin{align}
\label{eq:chi}
\chi^2_L = 2\sum_i \bigg[ y_i - n_i + n_i \log \Big(\frac{n_i}{y_i}\Big)\bigg] ,
\end{align}
where for each data bin, $i$, $n_i$ is the number of observed counts and $y_i$ is the number of counts predicted by the model. This statistic has, at least, two advantages over the standard chi-square: It preserves the area of the spectrum and it correctly treats bins with only a few or zero counts. We then use the MINUIT2 minimisation toolbox~\cite{James1994} to minimise $\chi^2_L$, using the level energies, partial widths and $\beta$ feedings as fitting parameters. We include data in the range \SIrange{600}{2800}{\kilo\electronvolt}, since at \SI{600}{\kilo\electronvolt} we estimate the pure $\beta$ signal to be about two orders of magnitude below the $\alpha$ signal.

\begin{figure}[htbp]
\centering
\includegraphics[width=\columnwidth]{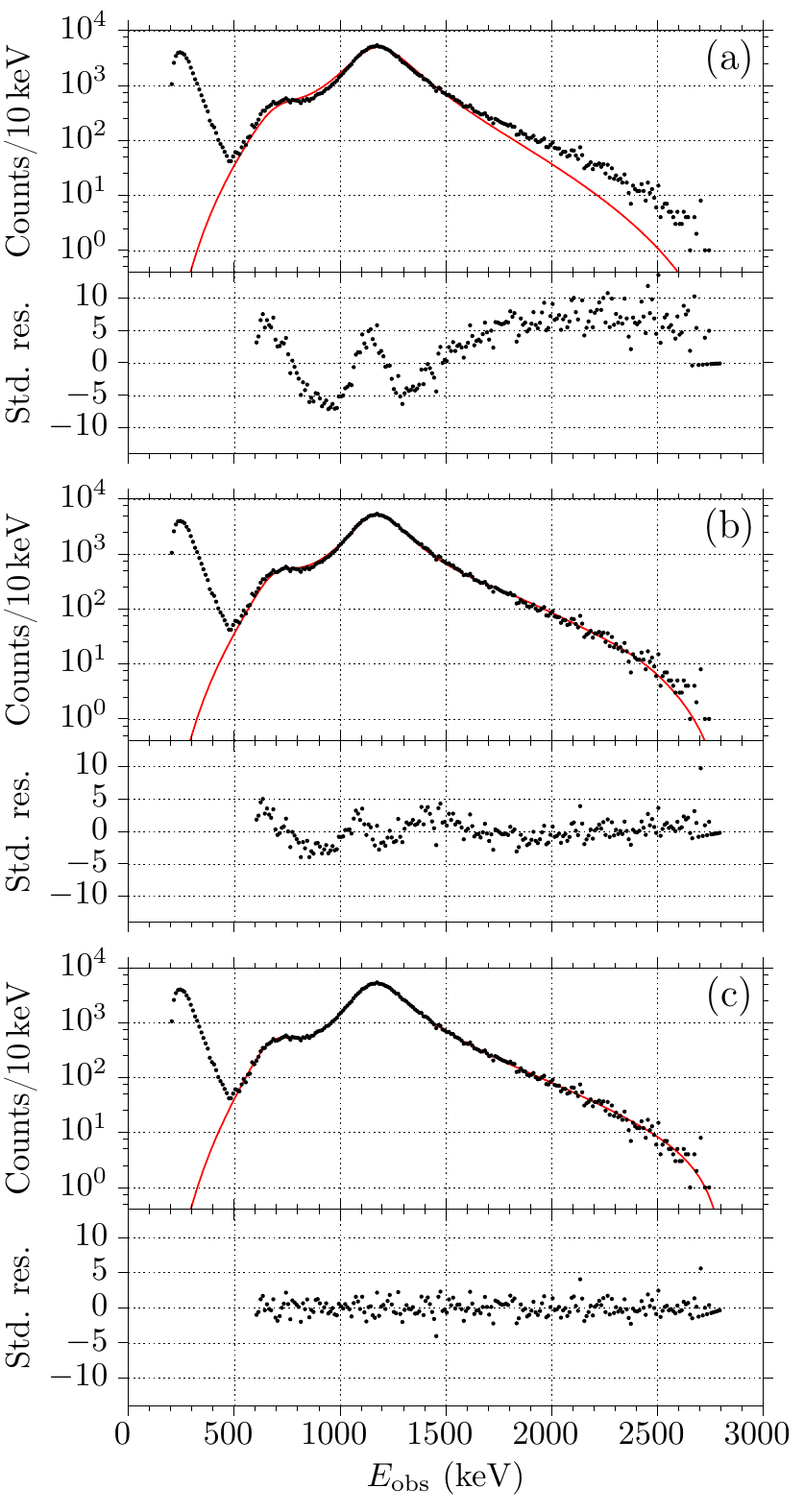}
\caption{The figure shows the observed decay spectrum together with the best $R$-matrix fits (solid red lines) for (a) model I, (b) model II and (c) model III. Also shown are the standardised residuals, $(n_i-y_i)/y_i$, corresponding to each fit (see the text following eq.~\eqref{eq:chi}).}
\label{fig:fit}
\end{figure}

\begin{table}[htbp]
\centering
\caption{Three models tested in the fit to the $\beta$-delayed $\alpha$ spectrum of ${^{11}\mathrm{Be}}$. Only channels where the $\alpha + {^7\mathrm{Li}}$ system has an orbital angular momentum of $l=1$ are included in the models.}
\label{tab:models}
\begin{ruledtabular}
\begin{tabular}{l c c c}
Model & I & II & III \\
\midrule
States & $3/2^+$ & $3/2^+ + 1/2^+$ & $3/2^+ + 3/2^+$ \\
$\chi^2_L/\mathrm{ndf}$ & \num{21.7} & \num{3.11} & \num{1.26} 
\end{tabular}
\end{ruledtabular}
\end{table}
In Ref.~\citep{Alburger1981} it was concluded that only the $3/2^+$ level at \SI{9.87}{\mega\electronvolt} played a role in the $\beta$-delayed $\alpha$ decay of $^{11}\mathrm{Be}$. In principle this state could emit $\alpha$ particles in both $l=1$ and $l=3$ channels, but since it sits not far above the threshold and still shows a considerable width, the spectrum must be dominated by $l=1$ emission. In a first attempt to fit the spectrum we therefore include only one $3/2^+$ level and $l=1$ channels (model I of \tref{tab:models}). The best fit is shown as the solid red line in \fref{fig:fit}(a).
It is immediately clear that this model does not fit the experimental data, since it lies above the experimental data on the low-energy side of the peak and below the experimental data on the high-energy side of the peak. Also the reduced chi-square value of $\chi^2_L / \mathrm{ndf} = 4710 / 217 \approx 21.7$ allows us reject this model. Allowing $l=3$ channels to contribute does not improve the fit, and we conclude that more than one level has to be involved in the decay.

Allowed transitions from the $1/2^+$ ground state of $^{11}\mathrm{Be}$ can only populate $1/2^+$ or $3/2^+$ states in $^{11}\mathrm{B}$. We therefore test two new models: One with an extra $1/2^+$ level (model II) and one with an extra $3/2^+$ level (model III) in addition to the well-known $3/2^+$ level at \SI{9.87}{\mega\electronvolt}. The fundamental difference between models II and III is that the decay amplitudes for two levels in model III must be added coherently, allowing for constructive and destructive interference in the spectrum. Fitting each of the models results in the reduced $\chi^2_L$ values listed in \tref{tab:models}. We see that, even though model II already improves the fit significantly, only model III provides an acceptable fit to our data. The best fit of the three models are shown in \fref{fig:fit}, together with the standardised residuals. We see that the effect of the interference is to suppress (enhance) the spectrum on the low- (high-) energy side of the main peak. The best fit parameters for model III are listed in \tref{tab:fits} for three different values of the channel radius.

\begin{table}[htbp]
\centering
\caption{Best fit parameters for three different values of the channel radius. The subscripts are the $\lambda$- or $\lambda c$-designation, where $c=1,2$ denote the ${^7\mathrm{Li(gs)}}+\alpha$ and ${^7\mathrm{Li(478)}}+\alpha$ channel, respectively. The quoted errors are the statistical errors provided by the MINUIT2 package. Parameters in square brackets are not used in the fit but are derived quantities. There are 217 degrees of freedom in the fits.}
\label{tab:fits}
\begin{ruledtabular}
%\begin{tabular}{p{0.25\columnwidth} p{0.23\columnwidth} p{0.23\columnwidth} p{0.17\columnwidth}}
\begin{tabular}{l c c c}
%\toprule
$r_0$ $(\si{\femto\meter})$ & 1.4 & 1.5 & 1.6 \\
$a_c$ $(\si{\femto\meter})$ & 4.90 & 5.25 & 5.60 \\
\midrule
$E_1$ $(\si{\kilo\electronvolt})$ & \num{9850 \pm 1} & \num{9848 \pm 1} & \num{9846 \pm 1} \\
$\Gamma_{11}$ $(\si{\kilo\electronvolt})$ & \num{240 \pm 3} & \num{237 \pm 3} & \num{233 \pm 3} \\
$\Gamma_{12}$ $(\si{\kilo\electronvolt})$ & \num{21.3 \pm 0.3} & \num{20.8 \pm 0.3} & \num{20.4 \pm 0.3} \\
$B_1 / \sqrt{N}$ & \num{0.265 \pm 0.005} & \num{0.190 \pm 0.002} & \num{0.161 \pm 0.002} \\
$[\theta_{11}^2]$ & \num{4.30 \pm 0.05} & \num{1.99 \pm 0.03} & \num{1.31 \pm 0.02} \\
$[\theta_{12}^2]$ & \num{3.19 \pm 0.05} & \num{1.38 \pm 0.02} & \num{0.84 \pm 0.02} \\
$[M_{GT,1}]$ & \num{0.726 \pm 0.018} & \num{0.722 \pm 0.013} & \num{0.717 \pm 0.012} \\
$[B_{GT,1}]$ & \num{0.326 \pm 0.016} & \num{0.322 \pm 0.012} & \num{0.318 \pm 0.011} \\
$[\log (ft)_1]$ & \num{4.067 \pm 0.049} & \num{4.072 \pm 0.036} & \num{4.078 \pm 0.033} \\
\midrule
$E_2$ $(\si{\kilo\electronvolt})$ & \num{11474 \pm 71} & \num{11475 \pm 75} & \num{11492 \pm 79} \\
$\Gamma_{21}$ $(\si{\kilo\electronvolt})$\footnotemark[1] & \num{-361 \pm 163} & \num{-388 \pm 157} & \num{-431 \pm 145} \\
$\Gamma_{22}$ $(\si{\kilo\electronvolt})$ & \num{76 \pm 71} & \num{66 \pm 68} & \num{47 \pm 59} \\
$B_2 / \sqrt{N}$ & \num{ 0.181\pm 0.034} & \num{0.166\pm 0.029} & \num{0.156 \pm 0.026} \\
$[\theta_{21}^2]$\footnotemark[1] & \num{-0.18 \pm 0.08} & \num{-0.19 \pm 0.08} & \num{-0.21 \pm 0.07} \\
$[\theta_{22}^2]$ & \num{0.049 \pm 0.045} & \num{0.041 \pm 0.043} & \num{0.029 \pm 0.037} \\
$[M_{GT,2}]$ & \num{1.21 \pm 0.23} & \num{1.11 \pm 0.19} & \num{1.05 \pm 0.17} \\
$[B_{GT,2}]$ & \num{0.90 \pm 0.35} & \num{0.77 \pm 0.26} & \num{0.67 \pm 0.22} \\
$[\log (ft)_2]$ & \num{3.63 \pm 0.39} & \num{3.70 \pm 0.34} & \num{3.75 \pm 0.33} \\
\midrule
$\mu a_c^2 / \hbar^2$ $(\si{\kilo\electronvolt})$ & \num{683.2} & \num{595.1} & \num{523.0} \\
$\chi^2_L$ & 269.42 &  269.30 & 269.33 
%\bottomrule
\end{tabular}
\footnotetext{The sign in these entries indicate the sign on the reduced width amplitude.}
\end{ruledtabular}
\end{table}

The $\beta$-decay feeding parameters, $B_\lambda$, do not have a clear, physical interpretation. As is apparent from \tref{tab:fits}, the feeding parameters are also sensitive to our choice of channel radius, which should not be the case for an observable parameter. Barker and Warburton~\cite{Barker1988} give approximate formulas that relate the $B_\lambda$'s to more useful quantities, for instance the Gamow-Teller matrix elements (see also Ref.~\cite{Warburton1986})
\begin{align}
M_{GT, \lambda} = \biggl ( \frac{\pi B}{N t_{\frac{1}{2}}} \biggr )^{\frac{1}{2}} \biggl ( 1 + \sum_c \gamma_{\lambda c}^2 \frac{d S_c}{dE}\bigg\vert_{E_\lambda} \bigg )^{-\frac{1}{2}} B_\lambda ,
\end{align}
where $B=\SI{6147 \pm 2}{\second}$~\cite{Hardy2005}, $N=\num{1.802e5}$ is the number of observed decays, $t_{\frac{1}{2}}=\SI{418 \pm 13}{\second}$ is the partial half-life of the $\beta$-delayed $\alpha$ decay (calculated using the value of $b_{\beta\alpha}$ found in Sec.~\ref{sec:branching}) and $S_c$ is the $R$-matrix shift function. It is also possible to derive approximate $B_{GT,\lambda}$ and $(ft)_\lambda$ values:
\begin{align}
B_{GT, \lambda} = \Big(\frac{g_A}{g_V}\Big)^{-2} M_{GT, \lambda} ^2 \qquad ; \; (ft)_\lambda = \frac{B}{M_{GT, \lambda}^2} ,
\end{align}
which may be more familiar measures of the $\beta$ transition strength (in the above formula $\lvert g_A / g_V \rvert = \num{1.2723 \pm 0.0023}$~\cite{Tanabashi2018}). These quantities are also included in \tref{tab:fits}, however, since we are dealing with broad, overlapping resonances, one must be careful when interpreting their values, see for instance the discussions in Refs.~\cite{Barker1988,Riisager2014}.

Also listed in \tref{tab:fits} are the dimensionless reduced widths, calculated using the definition from Lane and Thomas~\cite{Lane1958}:
\begin{align}
\theta_{\lambda c}^2 = \frac{\gamma_{\lambda c}^2 \mu a_c^2}{ \hbar^2} ,
\end{align}
where $\mu$ is the reduced mass of the ${^7\mathrm{Li}}+\alpha$ system. The dimensionless reduced widths are useful when we attempt to judge whether a particular level is broad or narrow in an absolute sense. An approximate maximum limit for $\theta_{\lambda c}^2$ is given by Teichmann and Wigner~\cite{Teichmann1952}: $\theta_{\lambda c}^2 < 1.5$. The fits performed with $r_0 = \SI{1.4}{\femto\meter}$ and \SI{1.5}{\femto\meter} result in reduced widths for the \SI{9.87}{\mega\electronvolt} level which are very large compared with this limit, suggesting that an unusually large channel radius is required to accomodate the observed width of this state. Based on these considerations we think that the result of the fit with $r_0=\SI{1.6}{\femto\meter}$ should be preferred over the results found using the smaller channel radii.

Finally, it is worthwhile to investigate the importance of possible systematic errors for our results. The most obvious source of systematic error is the energy calibration, since it was performed with an external calibration source. Calculation of energy losses through the detector dead layers is always connected with some uncertainty, to which must be added possible, unknown energy losses in the calibration source itself. We consider it reasonable to include a $\pm \SI{10}{\kilo\electronvolt}$ error in the energy calibration. Varying the energy calibration and detector resolution within the quoted uncertainties and refitting the spectrum we estimate the effect of systematic errors by observing the resulting variation in the best fit parameters. The results are listed in \tref{tab:results}, and it is reassuring to see that our analysis is quite robust to possible systematic calibration errors.
\begin{table}[htbp]
\centering
\caption{Observable parameters derived from the results of the $R$-matrix fitting to the $\beta$-delayed $\alpha$ spectrum of ${^{11}\mathrm{Be}}$. The statistical uncertainties are shown in round parantheses while possible errors due to systematics are quoted in square brackets.}
\label{tab:results}
\begin{ruledtabular}
%\begin{tabular}{p{0.3\columnwidth} p{0.25\columnwidth} p{0.25\columnwidth}}
\begin{tabular}{l c c}
 & $\lambda = 1$ & $\lambda = 2$ \\
\midrule
$E_\lambda$ $(\si{\kilo\electronvolt})$ & \num{9846 \pm 1}[10] & \num{11490 \pm 80}[50] \\
$\Gamma_{\lambda 1}$ $(\si{\kilo\electronvolt})$ & \num{233 \pm 3}[3] & \num{430 \pm 150}[50]  \\
$\Gamma_{\lambda 2}$ $(\si{\kilo\electronvolt})$ & \num{20.4 \pm 0.3}[3] & \num{50 \pm 60}[50] \\
$M_{GT,\lambda}$ & \num{0.717 \pm 0.012}[7] & \num{1.05 \pm 0.17}[5]  \\
$B_{GT,\lambda}$ & \num{0.318 \pm 0.011}[6] & \num{0.7 \pm 0.2}[1]  \\
$\log (ft)_\lambda$ & \num{4.08 \pm 0.03}[2] & \num{3.8 \pm 0.3}[1]  \\
\end{tabular}
\end{ruledtabular}
\end{table}

\section{Results \& Discussion}
We determined the branching ratio for $\beta$-delayed $\alpha$ decay to \SI{3.30 \pm 0.10}{\percent}. This value is in agreement with the earlier result \SI{3.1 \pm 0.4}{\percent}, found in Refs.~\cite{Alburger1981,Millener1982}, but slightly more precise.
\begin{figure}[htbp]
\centering
\includegraphics[width=\columnwidth]{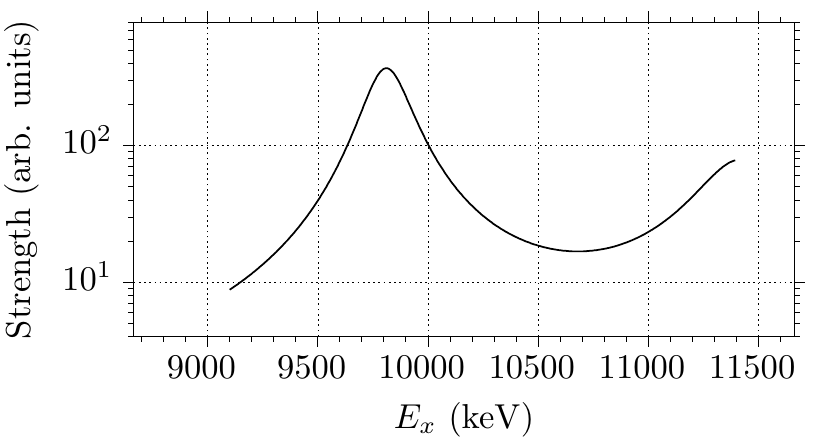}
\caption{The resonant strength in the $^{11}\mathrm{B}$ system as function of excitation energy. The graph is essentially equivalent to the delayed $\alpha$ spectrum in \fref{fig:fit}, but with the effects of $\beta$-decay phase space and the Coulomb barrier removed.}
\label{fig:strength}
\end{figure}
We obtained a satisfactory $R$-matrix fit to the observed spectrum by including two $3/2^+$ resonances in the ${^{11}\mathrm{B}}$ daughter system and only $p$-wave alphas in the outgoing channels. The derived resonant strength is shown in \fref{fig:strength}, and we clearly see the upturn at high breakup energies, confirming the tentative observation by Alburger \textit{et al.}~\cite{Alburger1981}. By integrating the contributions from the ${^7\mathrm{Li(gs)}} + \alpha$ and ${^7\mathrm{Li(478)}} + \alpha$ channels over the $Q_{\beta}$ window we find the branching ratios to be \SI{92.1 \pm 0.3}{\percent} and \SI{7.9 \pm 0.3}{\percent}, respectively. These figures are not consistent with the currently accepted values of \SI{87.4 \pm 1.2}{\percent} and \SI{12.6 \pm 1.2}{\percent}~\cite{Kelley2012,Alburger1981}. Furthermore, from our results we deduce the intensity of the \SI{478}{\kilo\electronvolt} $\gamma$ line following the ${^{11}\mathrm{Be}}$ decay to be $I_{478} = \SI{0.261 \pm 0.013}{\percent}$ and, using the litterature value for the intensity of the \SI{2124}{\kilo\electronvolt} $\gamma$ line of $I_{2124} = \SI{35.5 \pm 1.8}{\percent}$~\cite{Kelley2012} (this number includes feeding from higher-lying states in ${^{11}\mathrm{B}}$), we obtain the relative intensity of the two $\gamma$ lines $I_{478}/I_{2124} = \SI{0.74 \pm 0.05}{\percent}$.
%From literature values alone the ratio instead becomes $I_{478}/I_{2124} = \SI{1.10 \pm 0.19}{\percent}$.
In a recent experiment, the ratio was measured to be $I_{478}/I_{2124} = \SI{0.75 \pm 0.06}{\percent}$~\cite{Borge2013}, in excellent agreement with our result. The new determination of $b_{\beta\alpha}$ and $I_{478}$ provides two, independent ways of normalising the ${^{11}\mathrm{Be}}$ decay as alternatives to measuring the intensity of the \SI{2124}{\kilo\electronvolt} $\gamma$ line, namely by counting the emitted alphas and/or the \SI{478}{\kilo\electronvolt} gammas. A reliable normalisation of this decay is important, for instance for constraining the branching ratio of ${^{11}\mathrm{Li}}$ decaying to the ground state of ${^{11}\mathrm{Be}}$~\cite{Borge1997}, as well as for the characterisation of the rare $\beta$-delayed proton decay of ${^{11}\mathrm{Be}}$~\cite{Riisager2014b}.

From \fref{fig:strength} it is clear that the spectrum cannot be explained by a single contribution from the \SI{9.87}{\mega\electronvolt}-state, but that some extra strength is needed. We interpret the increase of the spectrum above $E_x \sim \SI{10.7}{\mega\electronvolt}$ as the low-energy tail of a broad level with its centroid energy above the $Q_\beta$ window. From the interference pattern we conclude that this level has to have $3/2^+$ symmetry. The high-energy strength can be a sign of either a resonance in ${^{11}\mathrm{B}}$ or of decays proceeding directly to the continuum. Direct decays are usually incorporated in the $R$-matrix framework by the introduction of a so-called background level, however, such a level most often appears with unrealistic level parameters, for instance extremely large width or $\beta$ feeding, and the best fit parameters tend to depend strongly on the choice of channel radius~\cite{Riisager2015}. From the results in \tref{tab:fits} the parameters of the second level appear quite robust to changes in channel radius, and the values are not unphysically large. From these considerations it seems most reasonable to interpret this second $R$-matrix level as a real state in ${^{11}\mathrm{B}}$. There are presently no known $3/2^+$ states at $E_x = \SI{11.49}{\mega\electronvolt}$ in ${^{11}\mathrm{B}}$, but there is a state in the evaluation at \SI{11.450 \pm 0.017}{\mega\electronvolt} and a width of \SI{93 \pm 17}{\kilo\electronvolt} with no $J^\pi$ assignment~\cite{Kelley2012}. This state has been observed in several experiments~\cite{Bichsel1957,Cusson1966,Zwieglinski1982}, and in particular it has been observed in elastic $\alpha$ scattering, which should indeed be the case for a state with such a considerable $\alpha$ width. Based on our results we believe it is reasonable to assign $J^\pi = 3/2^+$ to this state.

It is worth noting that the level energy we find for the \SI{9.87}{\mega\electronvolt} state is a few tens of \si{\kilo\electronvolt} lower than the tabulated value of \SI{9873 \pm 4}{\kilo\electronvolt}~\cite{Kelley2012}. This discrepancy must either be explained by a serious systematic error in our analysis or by a real shift in the observed peak position due to the nature of the $\beta$ transition (the evaluation is based primarily on scattering and reaction experiments). It has recently been pointed out that the energy dependence of the $\beta$-decay matrix element can cause a distortion of delayed particle spectra which shifts the observed peak towards lower energy~\cite{Riisager2015}. The effect is most pronounced for decays through very broad levels, and in the present case it is not unrealistic that it could explain a shift of the observed magnitude. While the level energy deviates slightly from the evaluation, the observed width is more than a factor of two larger than the tabulated value of \SI{109 \pm 14}{\kilo\electronvolt} (again, a result based on scattering and reaction experiments). This result confirms the observation made by Alburger \textit{et al.}~\citep{Alburger1981}. We are not aware of any mechanism that can explain such a strong dependence of the width on the reaction mechanism, and it could indeed be interesting to have this width re-measured in alternative channels in order to investigate this effect in more detail.

%Also from an $R$-matrix perspective the \SI{9.87}{\mega\electronvolt} level is very broad, and we found it necessary to use a quite large channel radius 
The \SI{9.87}{\mega\electronvolt} level is also very broad in an $R$-matrix context, and we found it necessary to use a quite large channel radius in our model in order to obey the Wigner limit. The large channel radius indicates that the wave function is spatially extended, and, based on the large $\alpha$ width, we can speculate that this state has a well-developed ${^7\mathrm{Li}} + \alpha$ cluster structure. The \SI{11.49}{\mega\electronvolt} level appears with a more moderate $\alpha$ width, suggesting that $\alpha$ clusterisation could be important, but is probably not the dominant feature of this state.

Both levels have quite large $B_{GT}$ values, summing approximately to 1. The sum rule for GT strength gives an upper limit of $3(N-Z) = 9$, and since the total GT strength to the bound levels in ${^{11}\mathrm{B}}$ is only approximately \num{0.016}, our result for the two $3/2^+$ levels does not violate this limit. In fact, a major part of the GT strength is expected to appear at energies close to the mother state~\cite{Sagawa1993}, so also from that perspective our results are sensible. It might seem somewhat counter-intuitive that the ${^{11}\mathrm{Be}}$ ground state, which has a ${^{10}\mathrm{Be}}$ core with a surrounding neutron halo, should have such a considerable overlap with $\alpha$-clusterised states in ${^{11}\mathrm{B}}$. AMD calculations suggest, however, that ${^{10}\mathrm{Be}}$ is itself clusterised with an appreciable $\alpha + \alpha + 2n$ component~\cite{KanadaEnyo1995}, and it is not difficult to imagine that the $\alpha$ clusters could to some extent survive the rearrangement caused by the $\beta$ transition.

That an extra $3/2^+$ state should appear in ${^{11}\mathrm{B}}$ is not entirely surprising. In a shell-model calculation Teeters and Kurath~\cite{Teeters1977} attempted to calculate the positive-parity states in ${^{11}\mathrm{B}}$, and they found four $3/2^+$ states below \SI{13}{\mega\electronvolt}, while in the current evaluation there are only two $3/2^+$ states identified below \SI{13}{\mega\electronvolt}, so there are a couple of $3/2^+$ states missing. Another feature of the shell-model calculation is that, while most of the calculated states agree very well with the experimentally determined states, the calculated $3/2^+$ states appear almost \SI{1.5}{\mega\electronvolt} higher than the experimental states. Clusterised states are notoriously difficult to obtain from shell model calculations, the Hoyle state in the neighbouring ${^{12}\mathrm{C}}$ system being another prime example~\cite{Cohen1965}, and if we accept the hypothesis that the $3/2^+$ states in ${^{11}\mathrm{B}}$ contain large cluster components, the discrepancy can to some extend be understood. Taking into consideration that the calculations by Teeters and Kurath~\cite{Teeters1977} were performed more than 40 years ago, it would be interesting to see a more modern calculation of the positive parity states in ${^{11}\mathrm{B}}$, and in particular to have theoretical estimates of the GT matrix elements with ${^{11}\mathrm{Be}}$.

\section{Conclusion}
We have measured the $\beta$-delayed $\alpha$ spectrum from ${^{11}\mathrm{Be}}$. Based on our analysis of the data we have determined the absolute branching ratios for delayed ${^7\mathrm{Li(gs)}} + \alpha$ and ${^7\mathrm{Li(478)}} + \alpha$ breakup. Using the $R$-matrix framework we concluded that two $3/2^+$ levels contribute to the decay, the known state at \SI{9.87}{\mega\electronvolt} and a second state at \SI{11.49 \pm 0.1}{\mega\electronvolt}. The extremely large observed width of the former state requires the use of a large channel radius, indicative of an extended wave function with a large cluster component. Based on its energy and width we propose to identify the second state with an already observed state at \SI{11.45}{\mega\electronvolt} which has no spin or parity assignment. The two states in our analysis both have significant $B_{GT}$'s, which points towards large overlaps between these states and the ${^{11}\mathrm{Be}}$ ground state. It is left as an open question why the observed width of the \SI{9.87}{\mega\electronvolt} state appears approximately a factor of two larger in $\beta$-decay experiments compared to scattering and reaction experiments. We believe this discrepancy warrants a remeasurement of this quantity in alternative channels.

\section{Acknowledgements}
\noindent We acknowledge the support of the ISOLDE Collaboration and technical teams. This work has been funded by FWO-Vlaanderen (Belgium), by BOF KU Leuven (GOA/2010/010), by the Interuniversity Attraction Poles Programme initiated by the Belgian Science Policy Office (BriX network P7/12) and by the European Union’s Seventh Framework Programme for Research and Technological Development under grant agreement No 262010. 

\newpage
\bibliography{bibliography}

%merlin.mbs apsrev4-1.bst 2010-07-25 4.21a (PWD, AO, DPC) hacked
%Control: key (0)
%Control: author (8) initials jnrlst
%Control: editor formatted (1) identically to author
%Control: production of article title (-1) disabled
%Control: page (0) single
%Control: year (1) truncated
%Control: production of eprint (0) enabled
\begin{thebibliography}{51}%
\makeatletter
\providecommand \@ifxundefined [1]{%
 \@ifx{#1\undefined}
}%
\providecommand \@ifnum [1]{%
 \ifnum #1\expandafter \@firstoftwo
 \else \expandafter \@secondoftwo
 \fi
}%
\providecommand \@ifx [1]{%
 \ifx #1\expandafter \@firstoftwo
 \else \expandafter \@secondoftwo
 \fi
}%
\providecommand \natexlab [1]{#1}%
\providecommand \enquote  [1]{``#1''}%
\providecommand \bibnamefont  [1]{#1}%
\providecommand \bibfnamefont [1]{#1}%
\providecommand \citenamefont [1]{#1}%
\providecommand \href@noop [0]{\@secondoftwo}%
\providecommand \href [0]{\begingroup \@sanitize@url \@href}%
\providecommand \@href[1]{\@@startlink{#1}\@@href}%
\providecommand \@@href[1]{\endgroup#1\@@endlink}%
\providecommand \@sanitize@url [0]{\catcode `\\12\catcode `\$12\catcode
  `\&12\catcode `\#12\catcode `\^12\catcode `\_12\catcode `\%12\relax}%
\providecommand \@@startlink[1]{}%
\providecommand \@@endlink[0]{}%
\providecommand \url  [0]{\begingroup\@sanitize@url \@url }%
\providecommand \@url [1]{\endgroup\@href {#1}{\urlprefix }}%
\providecommand \urlprefix  [0]{URL }%
\providecommand \Eprint [0]{\href }%
\providecommand \doibase [0]{http://dx.doi.org/}%
\providecommand \selectlanguage [0]{\@gobble}%
\providecommand \bibinfo  [0]{\@secondoftwo}%
\providecommand \bibfield  [0]{\@secondoftwo}%
\providecommand \translation [1]{[#1]}%
\providecommand \BibitemOpen [0]{}%
\providecommand \bibitemStop [0]{}%
\providecommand \bibitemNoStop [0]{.\EOS\space}%
\providecommand \EOS [0]{\spacefactor3000\relax}%
\providecommand \BibitemShut  [1]{\csname bibitem#1\endcsname}%
\let\auto@bib@innerbib\@empty
%</preamble>
\bibitem [{\citenamefont {Riisager}(2013)}]{Riisager2013}%
  \BibitemOpen
  \bibfield  {author} {\bibinfo {author} {\bibfnamefont {K.}~\bibnamefont
  {Riisager}},\ }\href {http://stacks.iop.org/1402-4896/2013/i=T152/a=014001}
  {\bibfield  {journal} {\bibinfo  {journal} {Physica Scripta}\ }\textbf
  {\bibinfo {volume} {2013}},\ \bibinfo {pages} {014001} (\bibinfo {year}
  {2013})}\BibitemShut {NoStop}%
\bibitem [{\citenamefont {Kelley}\ \emph {et~al.}(2012)\citenamefont {Kelley},
  \citenamefont {Kwan}, \citenamefont {Purcell}, \citenamefont {Sheu},\ and\
  \citenamefont {Weller}}]{Kelley2012}%
  \BibitemOpen
  \bibfield  {author} {\bibinfo {author} {\bibfnamefont {J.}~\bibnamefont
  {Kelley}}, \bibinfo {author} {\bibfnamefont {E.}~\bibnamefont {Kwan}},
  \bibinfo {author} {\bibfnamefont {J.}~\bibnamefont {Purcell}}, \bibinfo
  {author} {\bibfnamefont {C.}~\bibnamefont {Sheu}}, \ and\ \bibinfo {author}
  {\bibfnamefont {H.}~\bibnamefont {Weller}},\ }\href {\doibase
  https://doi.org/10.1016/j.nuclphysa.2012.01.010} {\bibfield  {journal}
  {\bibinfo  {journal} {Nuclear Physics A}\ }\textbf {\bibinfo {volume}
  {880}},\ \bibinfo {pages} {88 } (\bibinfo {year} {2012})}\BibitemShut
  {NoStop}%
\bibitem [{\citenamefont {Alburger}\ and\ \citenamefont
  {Engelbertink}(1970)}]{Alburger1970}%
  \BibitemOpen
  \bibfield  {author} {\bibinfo {author} {\bibfnamefont {D.~E.}\ \bibnamefont
  {Alburger}}\ and\ \bibinfo {author} {\bibfnamefont {G.~A.~P.}\ \bibnamefont
  {Engelbertink}},\ }\href {\doibase 10.1103/PhysRevC.2.1594} {\bibfield
  {journal} {\bibinfo  {journal} {Phys. Rev. C}\ }\textbf {\bibinfo {volume}
  {2}},\ \bibinfo {pages} {1594} (\bibinfo {year} {1970})}\BibitemShut
  {NoStop}%
\bibitem [{\citenamefont {Riisager}\ \emph {et~al.}(2014)\citenamefont
  {Riisager} \emph {et~al.}}]{Riisager2014b}%
  \BibitemOpen
  \bibfield  {author} {\bibinfo {author} {\bibfnamefont {K.}~\bibnamefont
  {Riisager}} \emph {et~al.},\ }\href {\doibase
  https://doi.org/10.1016/j.physletb.2014.03.062} {\bibfield  {journal}
  {\bibinfo  {journal} {Physics Letters B}\ }\textbf {\bibinfo {volume}
  {732}},\ \bibinfo {pages} {305 } (\bibinfo {year} {2014})}\BibitemShut
  {NoStop}%
\bibitem [{\citenamefont {Fornal}\ and\ \citenamefont
  {Grinstein}(2018)}]{Fornal2018}%
  \BibitemOpen
  \bibfield  {author} {\bibinfo {author} {\bibfnamefont {B.}~\bibnamefont
  {Fornal}}\ and\ \bibinfo {author} {\bibfnamefont {B.}~\bibnamefont
  {Grinstein}},\ }\href {\doibase 10.1103/PhysRevLett.120.191801} {\bibfield
  {journal} {\bibinfo  {journal} {Phys. Rev. Lett.}\ }\textbf {\bibinfo
  {volume} {120}},\ \bibinfo {pages} {191801} (\bibinfo {year}
  {2018})}\BibitemShut {NoStop}%
\bibitem [{\citenamefont {Pf\"utzner}\ and\ \citenamefont
  {Riisager}(2018)}]{Pfutzner2018}%
  \BibitemOpen
  \bibfield  {author} {\bibinfo {author} {\bibfnamefont {M.}~\bibnamefont
  {Pf\"utzner}}\ and\ \bibinfo {author} {\bibfnamefont {K.}~\bibnamefont
  {Riisager}},\ }\href {\doibase 10.1103/PhysRevC.97.042501} {\bibfield
  {journal} {\bibinfo  {journal} {Phys. Rev. C}\ }\textbf {\bibinfo {volume}
  {97}},\ \bibinfo {pages} {042501} (\bibinfo {year} {2018})}\BibitemShut
  {NoStop}%
\bibitem [{\citenamefont {Alburger}\ \emph {et~al.}(1981)\citenamefont
  {Alburger}, \citenamefont {Millener},\ and\ \citenamefont
  {Wilkinson}}]{Alburger1981}%
  \BibitemOpen
  \bibfield  {author} {\bibinfo {author} {\bibfnamefont {D.~E.}\ \bibnamefont
  {Alburger}}, \bibinfo {author} {\bibfnamefont {D.~J.}\ \bibnamefont
  {Millener}}, \ and\ \bibinfo {author} {\bibfnamefont {D.~H.}\ \bibnamefont
  {Wilkinson}},\ }\href {\doibase 10.1103/PhysRevC.23.473} {\bibfield
  {journal} {\bibinfo  {journal} {Phys. Rev. C}\ }\textbf {\bibinfo {volume}
  {23}},\ \bibinfo {pages} {473} (\bibinfo {year} {1981})}\BibitemShut
  {NoStop}%
\bibitem [{\citenamefont {Millener}\ \emph {et~al.}(1982)\citenamefont
  {Millener}, \citenamefont {Alburger}, \citenamefont {Warburton},\ and\
  \citenamefont {Wilkinson}}]{Millener1982}%
  \BibitemOpen
  \bibfield  {author} {\bibinfo {author} {\bibfnamefont {D.~J.}\ \bibnamefont
  {Millener}}, \bibinfo {author} {\bibfnamefont {D.~E.}\ \bibnamefont
  {Alburger}}, \bibinfo {author} {\bibfnamefont {E.~K.}\ \bibnamefont
  {Warburton}}, \ and\ \bibinfo {author} {\bibfnamefont {D.~H.}\ \bibnamefont
  {Wilkinson}},\ }\href {\doibase 10.1103/PhysRevC.26.1167} {\bibfield
  {journal} {\bibinfo  {journal} {Phys. Rev. C}\ }\textbf {\bibinfo {volume}
  {26}},\ \bibinfo {pages} {1167} (\bibinfo {year} {1982})}\BibitemShut
  {NoStop}%
\bibitem [{\citenamefont {Li}\ and\ \citenamefont {Sherr}(1954)}]{Li1954}%
  \BibitemOpen
  \bibfield  {author} {\bibinfo {author} {\bibfnamefont {C.~W.}\ \bibnamefont
  {Li}}\ and\ \bibinfo {author} {\bibfnamefont {R.}~\bibnamefont {Sherr}},\
  }\href {\doibase 10.1103/PhysRev.96.389} {\bibfield  {journal} {\bibinfo
  {journal} {Phys. Rev.}\ }\textbf {\bibinfo {volume} {96}},\ \bibinfo {pages}
  {389} (\bibinfo {year} {1954})}\BibitemShut {NoStop}%
\bibitem [{\citenamefont {Bichsel}\ and\ \citenamefont
  {Bonner}(1957)}]{Bichsel1957}%
  \BibitemOpen
  \bibfield  {author} {\bibinfo {author} {\bibfnamefont {H.}~\bibnamefont
  {Bichsel}}\ and\ \bibinfo {author} {\bibfnamefont {T.~W.}\ \bibnamefont
  {Bonner}},\ }\href {\doibase 10.1103/PhysRev.108.1025} {\bibfield  {journal}
  {\bibinfo  {journal} {Phys. Rev.}\ }\textbf {\bibinfo {volume} {108}},\
  \bibinfo {pages} {1025} (\bibinfo {year} {1957})}\BibitemShut {NoStop}%
\bibitem [{\citenamefont {Hinds}\ and\ \citenamefont
  {Middleton}(1959)}]{Hinds1959}%
  \BibitemOpen
  \bibfield  {author} {\bibinfo {author} {\bibfnamefont {S.}~\bibnamefont
  {Hinds}}\ and\ \bibinfo {author} {\bibfnamefont {R.}~\bibnamefont
  {Middleton}},\ }\href {http://stacks.iop.org/0370-1328/74/i=2/a=307}
  {\bibfield  {journal} {\bibinfo  {journal} {Proceedings of the Physical
  Society}\ }\textbf {\bibinfo {volume} {74}},\ \bibinfo {pages} {196}
  (\bibinfo {year} {1959})}\BibitemShut {NoStop}%
\bibitem [{\citenamefont {Cusson}(1966)}]{Cusson1966}%
  \BibitemOpen
  \bibfield  {author} {\bibinfo {author} {\bibfnamefont {R.}~\bibnamefont
  {Cusson}},\ }\href {\doibase https://doi.org/10.1016/0029-5582(66)90492-5}
  {\bibfield  {journal} {\bibinfo  {journal} {Nuclear Physics}\ }\textbf
  {\bibinfo {volume} {86}},\ \bibinfo {pages} {481 } (\bibinfo {year}
  {1966})}\BibitemShut {NoStop}%
\bibitem [{\citenamefont {Zwiegliński}\ \emph {et~al.}(1982)\citenamefont
  {Zwiegliński}, \citenamefont {Benenson}, \citenamefont {Crawley},
  \citenamefont {Galès},\ and\ \citenamefont {Weber}}]{Zwieglinski1982}%
  \BibitemOpen
  \bibfield  {author} {\bibinfo {author} {\bibfnamefont {B.}~\bibnamefont
  {Zwiegliński}}, \bibinfo {author} {\bibfnamefont {W.}~\bibnamefont
  {Benenson}}, \bibinfo {author} {\bibfnamefont {G.}~\bibnamefont {Crawley}},
  \bibinfo {author} {\bibfnamefont {S.}~\bibnamefont {Galès}}, \ and\ \bibinfo
  {author} {\bibfnamefont {D.}~\bibnamefont {Weber}},\ }\href {\doibase
  https://doi.org/10.1016/0375-9474(82)90521-8} {\bibfield  {journal} {\bibinfo
   {journal} {Nuclear Physics A}\ }\textbf {\bibinfo {volume} {389}},\ \bibinfo
  {pages} {301 } (\bibinfo {year} {1982})}\BibitemShut {NoStop}%
\bibitem [{\citenamefont {Smirnov}\ \emph {et~al.}(2005)\citenamefont
  {Smirnov}, \citenamefont {Aksouh}, \citenamefont {Dean}, \citenamefont
  {Witte}, \citenamefont {Huyse}, \citenamefont {Ivanov}, \citenamefont
  {Mayet}, \citenamefont {Mukha}, \citenamefont {Raabe}, \citenamefont
  {Thomas}, \citenamefont {Duppen}, \citenamefont {Angulo}, \citenamefont
  {Cabrera}, \citenamefont {Ninane},\ and\ \citenamefont
  {Davinson}}]{Smirnov2005}%
  \BibitemOpen
  \bibfield  {author} {\bibinfo {author} {\bibfnamefont {D.}~\bibnamefont
  {Smirnov}}, \bibinfo {author} {\bibfnamefont {F.}~\bibnamefont {Aksouh}},
  \bibinfo {author} {\bibfnamefont {S.}~\bibnamefont {Dean}}, \bibinfo {author}
  {\bibfnamefont {H.~D.}\ \bibnamefont {Witte}}, \bibinfo {author}
  {\bibfnamefont {M.}~\bibnamefont {Huyse}}, \bibinfo {author} {\bibfnamefont
  {O.}~\bibnamefont {Ivanov}}, \bibinfo {author} {\bibfnamefont
  {P.}~\bibnamefont {Mayet}}, \bibinfo {author} {\bibfnamefont
  {I.}~\bibnamefont {Mukha}}, \bibinfo {author} {\bibfnamefont
  {R.}~\bibnamefont {Raabe}}, \bibinfo {author} {\bibfnamefont {J.-C.}\
  \bibnamefont {Thomas}}, \bibinfo {author} {\bibfnamefont {P.~V.}\
  \bibnamefont {Duppen}}, \bibinfo {author} {\bibfnamefont {C.}~\bibnamefont
  {Angulo}}, \bibinfo {author} {\bibfnamefont {J.}~\bibnamefont {Cabrera}},
  \bibinfo {author} {\bibfnamefont {A.}~\bibnamefont {Ninane}}, \ and\ \bibinfo
  {author} {\bibfnamefont {T.}~\bibnamefont {Davinson}},\ }\href {\doibase
  https://doi.org/10.1016/j.nima.2005.03.165} {\bibfield  {journal} {\bibinfo
  {journal} {Nuclear Instruments and Methods in Physics Research Section A:
  Accelerators, Spectrometers, Detectors and Associated Equipment}\ }\textbf
  {\bibinfo {volume} {547}},\ \bibinfo {pages} {480 } (\bibinfo {year}
  {2005})}\BibitemShut {NoStop}%
\bibitem [{\citenamefont {Raabe}\ \emph {et~al.}(2009)\citenamefont {Raabe}
  \emph {et~al.}}]{Raabe2009}%
  \BibitemOpen
  \bibfield  {author} {\bibinfo {author} {\bibfnamefont {R.}~\bibnamefont
  {Raabe}} \emph {et~al.},\ }\href {\doibase 10.1103/PhysRevC.80.054307}
  {\bibfield  {journal} {\bibinfo  {journal} {Phys. Rev. C}\ }\textbf {\bibinfo
  {volume} {80}},\ \bibinfo {pages} {054307} (\bibinfo {year}
  {2009})}\BibitemShut {NoStop}%
\bibitem [{\citenamefont {Raabe}\ \emph {et~al.}(2008)\citenamefont {Raabe}
  \emph {et~al.}}]{Raabe2008}%
  \BibitemOpen
  \bibfield  {author} {\bibinfo {author} {\bibfnamefont {R.}~\bibnamefont
  {Raabe}} \emph {et~al.},\ }\href {\doibase 10.1103/PhysRevLett.101.212501}
  {\bibfield  {journal} {\bibinfo  {journal} {Phys. Rev. Lett.}\ }\textbf
  {\bibinfo {volume} {101}},\ \bibinfo {pages} {212501} (\bibinfo {year}
  {2008})}\BibitemShut {NoStop}%
\bibitem [{\citenamefont {Hyldegaard}\ \emph {et~al.}(2009)\citenamefont
  {Hyldegaard} \emph {et~al.}}]{Hyldegaard2009}%
  \BibitemOpen
  \bibfield  {author} {\bibinfo {author} {\bibfnamefont {S.}~\bibnamefont
  {Hyldegaard}} \emph {et~al.},\ }\href {\doibase
  https://doi.org/10.1016/j.physletb.2009.06.064} {\bibfield  {journal}
  {\bibinfo  {journal} {Physics Letters B}\ }\textbf {\bibinfo {volume}
  {678}},\ \bibinfo {pages} {459 } (\bibinfo {year} {2009})}\BibitemShut
  {NoStop}%
\bibitem [{\citenamefont {Roger}\ \emph {et~al.}(2012)\citenamefont {Roger}
  \emph {et~al.}}]{Roger2012}%
  \BibitemOpen
  \bibfield  {author} {\bibinfo {author} {\bibfnamefont {T.}~\bibnamefont
  {Roger}} \emph {et~al.},\ }\href {\doibase 10.1103/PhysRevLett.108.162502}
  {\bibfield  {journal} {\bibinfo  {journal} {Phys. Rev. Lett.}\ }\textbf
  {\bibinfo {volume} {108}},\ \bibinfo {pages} {162502} (\bibinfo {year}
  {2012})}\BibitemShut {NoStop}%
\bibitem [{\citenamefont {Refsgaard}\ \emph {et~al.}(2016)\citenamefont
  {Refsgaard}, \citenamefont {Kirsebom}, \citenamefont {Dijck}, \citenamefont
  {Fynbo}, \citenamefont {Lund}, \citenamefont {Portela}, \citenamefont
  {Raabe}, \citenamefont {Randisi}, \citenamefont {Renzi}, \citenamefont
  {Sambi}, \citenamefont {Sytema}, \citenamefont {Willmann},\ and\
  \citenamefont {Wilschut}}]{Refsgaard2016}%
  \BibitemOpen
  \bibfield  {author} {\bibinfo {author} {\bibfnamefont {J.}~\bibnamefont
  {Refsgaard}}, \bibinfo {author} {\bibfnamefont {O.}~\bibnamefont {Kirsebom}},
  \bibinfo {author} {\bibfnamefont {E.}~\bibnamefont {Dijck}}, \bibinfo
  {author} {\bibfnamefont {H.}~\bibnamefont {Fynbo}}, \bibinfo {author}
  {\bibfnamefont {M.}~\bibnamefont {Lund}}, \bibinfo {author} {\bibfnamefont
  {M.}~\bibnamefont {Portela}}, \bibinfo {author} {\bibfnamefont
  {R.}~\bibnamefont {Raabe}}, \bibinfo {author} {\bibfnamefont
  {G.}~\bibnamefont {Randisi}}, \bibinfo {author} {\bibfnamefont
  {F.}~\bibnamefont {Renzi}}, \bibinfo {author} {\bibfnamefont
  {S.}~\bibnamefont {Sambi}}, \bibinfo {author} {\bibfnamefont
  {A.}~\bibnamefont {Sytema}}, \bibinfo {author} {\bibfnamefont
  {L.}~\bibnamefont {Willmann}}, \ and\ \bibinfo {author} {\bibfnamefont
  {H.}~\bibnamefont {Wilschut}},\ }\href {\doibase
  https://doi.org/10.1016/j.physletb.2015.11.047} {\bibfield  {journal}
  {\bibinfo  {journal} {Physics Letters B}\ }\textbf {\bibinfo {volume}
  {752}},\ \bibinfo {pages} {296 } (\bibinfo {year} {2016})}\BibitemShut
  {NoStop}%
\bibitem [{\citenamefont {Borge}\ and\ \citenamefont
  {Blaum}(2018)}]{Borge2018}%
  \BibitemOpen
  \bibfield  {author} {\bibinfo {author} {\bibfnamefont {M.~J.~G.}\
  \bibnamefont {Borge}}\ and\ \bibinfo {author} {\bibfnamefont
  {K.}~\bibnamefont {Blaum}},\ }\href
  {http://stacks.iop.org/0954-3899/45/i=1/a=010301} {\bibfield  {journal}
  {\bibinfo  {journal} {Journal of Physics G: Nuclear and Particle Physics}\
  }\textbf {\bibinfo {volume} {45}},\ \bibinfo {pages} {010301} (\bibinfo
  {year} {2018})}\BibitemShut {NoStop}%
\bibitem [{\citenamefont {Mishin}\ \emph {et~al.}(1993)\citenamefont {Mishin},
  \citenamefont {Fedoseyev}, \citenamefont {Kluge}, \citenamefont {Letokhov},
  \citenamefont {Ravn}, \citenamefont {Scheerer}, \citenamefont {Shirakabe},
  \citenamefont {Sundell},\ and\ \citenamefont {Tengblad}}]{Mishin1993}%
  \BibitemOpen
  \bibfield  {author} {\bibinfo {author} {\bibfnamefont {V.}~\bibnamefont
  {Mishin}}, \bibinfo {author} {\bibfnamefont {V.}~\bibnamefont {Fedoseyev}},
  \bibinfo {author} {\bibfnamefont {H.-J.}\ \bibnamefont {Kluge}}, \bibinfo
  {author} {\bibfnamefont {V.}~\bibnamefont {Letokhov}}, \bibinfo {author}
  {\bibfnamefont {H.}~\bibnamefont {Ravn}}, \bibinfo {author} {\bibfnamefont
  {F.}~\bibnamefont {Scheerer}}, \bibinfo {author} {\bibfnamefont
  {Y.}~\bibnamefont {Shirakabe}}, \bibinfo {author} {\bibfnamefont
  {S.}~\bibnamefont {Sundell}}, \ and\ \bibinfo {author} {\bibfnamefont
  {O.}~\bibnamefont {Tengblad}},\ }\href {\doibase
  https://doi.org/10.1016/0168-583X(93)95839-W} {\bibfield  {journal} {\bibinfo
   {journal} {Nuclear Instruments and Methods in Physics Research Section B:
  Beam Interactions with Materials and Atoms}\ }\textbf {\bibinfo {volume}
  {73}},\ \bibinfo {pages} {550 } (\bibinfo {year} {1993})}\BibitemShut
  {NoStop}%
\bibitem [{\citenamefont {Kester}\ \emph {et~al.}(2003)\citenamefont {Kester}
  \emph {et~al.}}]{Kester2003}%
  \BibitemOpen
  \bibfield  {author} {\bibinfo {author} {\bibfnamefont {O.}~\bibnamefont
  {Kester}} \emph {et~al.},\ }\href {\doibase
  https://doi.org/10.1016/S0168-583X(02)01886-4} {\bibfield  {journal}
  {\bibinfo  {journal} {Nuclear Instruments and Methods in Physics Research
  Section B: Beam Interactions with Materials and Atoms}\ }\textbf {\bibinfo
  {volume} {204}},\ \bibinfo {pages} {20 } (\bibinfo {year} {2003})},\ \bibinfo
  {note} {14th International Conference on Electromagnetic Isotope Separators
  and Techniques Related to their Applications}\BibitemShut {NoStop}%
\bibitem [{\citenamefont {Sellin}\ \emph {et~al.}(1992)\citenamefont {Sellin},
  \citenamefont {Woods}, \citenamefont {Branford}, \citenamefont {Davinson},
  \citenamefont {Davis}, \citenamefont {Ireland}, \citenamefont {Livingston},
  \citenamefont {Page}, \citenamefont {Shotter}, \citenamefont {Hofmann},
  \citenamefont {Hunt}, \citenamefont {James}, \citenamefont {Hotchkis},
  \citenamefont {Freer},\ and\ \citenamefont {Thomas}}]{Sellin1992}%
  \BibitemOpen
  \bibfield  {author} {\bibinfo {author} {\bibfnamefont {P.}~\bibnamefont
  {Sellin}}, \bibinfo {author} {\bibfnamefont {P.}~\bibnamefont {Woods}},
  \bibinfo {author} {\bibfnamefont {D.}~\bibnamefont {Branford}}, \bibinfo
  {author} {\bibfnamefont {T.}~\bibnamefont {Davinson}}, \bibinfo {author}
  {\bibfnamefont {N.}~\bibnamefont {Davis}}, \bibinfo {author} {\bibfnamefont
  {D.}~\bibnamefont {Ireland}}, \bibinfo {author} {\bibfnamefont
  {K.}~\bibnamefont {Livingston}}, \bibinfo {author} {\bibfnamefont
  {R.}~\bibnamefont {Page}}, \bibinfo {author} {\bibfnamefont {A.}~\bibnamefont
  {Shotter}}, \bibinfo {author} {\bibfnamefont {S.}~\bibnamefont {Hofmann}},
  \bibinfo {author} {\bibfnamefont {R.}~\bibnamefont {Hunt}}, \bibinfo {author}
  {\bibfnamefont {A.}~\bibnamefont {James}}, \bibinfo {author} {\bibfnamefont
  {M.}~\bibnamefont {Hotchkis}}, \bibinfo {author} {\bibfnamefont
  {M.}~\bibnamefont {Freer}}, \ and\ \bibinfo {author} {\bibfnamefont
  {S.}~\bibnamefont {Thomas}},\ }\href {\doibase
  https://doi.org/10.1016/0168-9002(92)90867-4} {\bibfield  {journal} {\bibinfo
   {journal} {Nuclear Instruments and Methods in Physics Research Section A:
  Accelerators, Spectrometers, Detectors and Associated Equipment}\ }\textbf
  {\bibinfo {volume} {311}},\ \bibinfo {pages} {217 } (\bibinfo {year}
  {1992})}\BibitemShut {NoStop}%
\bibitem [{\citenamefont {Büscher}\ \emph {et~al.}(2008)\citenamefont
  {Büscher}, \citenamefont {Ponsaers}, \citenamefont {Raabe}, \citenamefont
  {Huyse}, \citenamefont {Duppen}, \citenamefont {Aksouh}, \citenamefont
  {Smirnov}, \citenamefont {Fynbo}, \citenamefont {Hyldegaard},\ and\
  \citenamefont {Diget}}]{Buscher2008}%
  \BibitemOpen
  \bibfield  {author} {\bibinfo {author} {\bibfnamefont {J.}~\bibnamefont
  {Büscher}}, \bibinfo {author} {\bibfnamefont {J.}~\bibnamefont {Ponsaers}},
  \bibinfo {author} {\bibfnamefont {R.}~\bibnamefont {Raabe}}, \bibinfo
  {author} {\bibfnamefont {M.}~\bibnamefont {Huyse}}, \bibinfo {author}
  {\bibfnamefont {P.~V.}\ \bibnamefont {Duppen}}, \bibinfo {author}
  {\bibfnamefont {F.}~\bibnamefont {Aksouh}}, \bibinfo {author} {\bibfnamefont
  {D.}~\bibnamefont {Smirnov}}, \bibinfo {author} {\bibfnamefont
  {H.}~\bibnamefont {Fynbo}}, \bibinfo {author} {\bibfnamefont
  {S.}~\bibnamefont {Hyldegaard}}, \ and\ \bibinfo {author} {\bibfnamefont
  {C.}~\bibnamefont {Diget}},\ }\href {\doibase
  https://doi.org/10.1016/j.nimb.2008.05.084} {\bibfield  {journal} {\bibinfo
  {journal} {Nuclear Instruments and Methods in Physics Research Section B:
  Beam Interactions with Materials and Atoms}\ }\textbf {\bibinfo {volume}
  {266}},\ \bibinfo {pages} {4652 } (\bibinfo {year} {2008})},\ \bibinfo {note}
  {proceedings of the XVth International Conference on Electromagnetic Isotope
  Separators and Techniques Related to their Applications}\BibitemShut
  {NoStop}%
\bibitem [{\citenamefont {Tarasov}\ and\ \citenamefont
  {Bazin}(2016)}]{Tarasov2016}%
  \BibitemOpen
  \bibfield  {author} {\bibinfo {author} {\bibfnamefont {O.}~\bibnamefont
  {Tarasov}}\ and\ \bibinfo {author} {\bibfnamefont {D.}~\bibnamefont
  {Bazin}},\ }\href {\doibase https://doi.org/10.1016/j.nimb.2016.03.021}
  {\bibfield  {journal} {\bibinfo  {journal} {Nuclear Instruments and Methods
  in Physics Research Section B: Beam Interactions with Materials and Atoms}\
  }\textbf {\bibinfo {volume} {376}},\ \bibinfo {pages} {185 } (\bibinfo {year}
  {2016})},\ \bibinfo {note} {proceedings of the XVIIth International
  Conference on Electromagnetic Isotope Separators and Related Topics
  (EMIS2015), Grand Rapids, MI, U.S.A., 11-15 May 2015}\BibitemShut {NoStop}%
\bibitem [{\citenamefont {Rytz}(1991)}]{Rytz1991}%
  \BibitemOpen
  \bibfield  {author} {\bibinfo {author} {\bibfnamefont {A.}~\bibnamefont
  {Rytz}},\ }\href {\doibase https://doi.org/10.1016/0092-640X(91)90002-L}
  {\bibfield  {journal} {\bibinfo  {journal} {Atomic Data and Nuclear Data
  Tables}\ }\textbf {\bibinfo {volume} {47}},\ \bibinfo {pages} {205 }
  (\bibinfo {year} {1991})}\BibitemShut {NoStop}%
\bibitem [{\citenamefont {Agostinelli}\ \emph {et~al.}(2003)\citenamefont
  {Agostinelli} \emph {et~al.}}]{Agostinelli2003}%
  \BibitemOpen
  \bibfield  {author} {\bibinfo {author} {\bibfnamefont {S.}~\bibnamefont
  {Agostinelli}} \emph {et~al.},\ }\href {\doibase
  https://doi.org/10.1016/S0168-9002(03)01368-8} {\bibfield  {journal}
  {\bibinfo  {journal} {Nuclear Instruments and Methods in Physics Research
  Section A: Accelerators, Spectrometers, Detectors and Associated Equipment}\
  }\textbf {\bibinfo {volume} {506}},\ \bibinfo {pages} {250 } (\bibinfo {year}
  {2003})}\BibitemShut {NoStop}%
\bibitem [{\citenamefont {Lennard}\ \emph {et~al.}(1986)\citenamefont
  {Lennard}, \citenamefont {Geissel}, \citenamefont {Winterbon}, \citenamefont
  {Phillips}, \citenamefont {Alexander},\ and\ \citenamefont
  {Forster}}]{Lennard1986}%
  \BibitemOpen
  \bibfield  {author} {\bibinfo {author} {\bibfnamefont {W.}~\bibnamefont
  {Lennard}}, \bibinfo {author} {\bibfnamefont {H.}~\bibnamefont {Geissel}},
  \bibinfo {author} {\bibfnamefont {K.}~\bibnamefont {Winterbon}}, \bibinfo
  {author} {\bibfnamefont {D.}~\bibnamefont {Phillips}}, \bibinfo {author}
  {\bibfnamefont {T.}~\bibnamefont {Alexander}}, \ and\ \bibinfo {author}
  {\bibfnamefont {J.}~\bibnamefont {Forster}},\ }\href {\doibase
  https://doi.org/10.1016/0168-9002(86)91033-8} {\bibfield  {journal} {\bibinfo
   {journal} {Nuclear Instruments and Methods in Physics Research Section A:
  Accelerators, Spectrometers, Detectors and Associated Equipment}\ }\textbf
  {\bibinfo {volume} {248}},\ \bibinfo {pages} {454 } (\bibinfo {year}
  {1986})}\BibitemShut {NoStop}%
\bibitem [{\citenamefont {Kirsebom}\ \emph {et~al.}(2014)\citenamefont
  {Kirsebom}, \citenamefont {Fynbo}, \citenamefont {Riisager}, \citenamefont
  {Raabe},\ and\ \citenamefont {Roger}}]{Kirsebom2014}%
  \BibitemOpen
  \bibfield  {author} {\bibinfo {author} {\bibfnamefont {O.}~\bibnamefont
  {Kirsebom}}, \bibinfo {author} {\bibfnamefont {H.}~\bibnamefont {Fynbo}},
  \bibinfo {author} {\bibfnamefont {K.}~\bibnamefont {Riisager}}, \bibinfo
  {author} {\bibfnamefont {R.}~\bibnamefont {Raabe}}, \ and\ \bibinfo {author}
  {\bibfnamefont {T.}~\bibnamefont {Roger}},\ }\href {\doibase
  https://doi.org/10.1016/j.nima.2014.05.005} {\bibfield  {journal} {\bibinfo
  {journal} {Nuclear Instruments and Methods in Physics Research Section A:
  Accelerators, Spectrometers, Detectors and Associated Equipment}\ }\textbf
  {\bibinfo {volume} {758}},\ \bibinfo {pages} {57 } (\bibinfo {year}
  {2014})}\BibitemShut {NoStop}%
\bibitem [{\citenamefont {Lane}\ and\ \citenamefont {Thomas}(1958)}]{Lane1958}%
  \BibitemOpen
  \bibfield  {author} {\bibinfo {author} {\bibfnamefont {A.~M.}\ \bibnamefont
  {Lane}}\ and\ \bibinfo {author} {\bibfnamefont {R.~G.}\ \bibnamefont
  {Thomas}},\ }\href {\doibase 10.1103/RevModPhys.30.257} {\bibfield  {journal}
  {\bibinfo  {journal} {Rev. Mod. Phys.}\ }\textbf {\bibinfo {volume} {30}},\
  \bibinfo {pages} {257} (\bibinfo {year} {1958})}\BibitemShut {NoStop}%
\bibitem [{\citenamefont {Vogt}(1962)}]{Vogt1962}%
  \BibitemOpen
  \bibfield  {author} {\bibinfo {author} {\bibfnamefont {E.}~\bibnamefont
  {Vogt}},\ }\href {\doibase 10.1103/RevModPhys.34.723} {\bibfield  {journal}
  {\bibinfo  {journal} {Rev. Mod. Phys.}\ }\textbf {\bibinfo {volume} {34}},\
  \bibinfo {pages} {723} (\bibinfo {year} {1962})}\BibitemShut {NoStop}%
\bibitem [{\citenamefont {Descouvemont}\ and\ \citenamefont
  {Baye}(2010)}]{Descouvemont2010}%
  \BibitemOpen
  \bibfield  {author} {\bibinfo {author} {\bibfnamefont {P.}~\bibnamefont
  {Descouvemont}}\ and\ \bibinfo {author} {\bibfnamefont {D.}~\bibnamefont
  {Baye}},\ }\href {http://stacks.iop.org/0034-4885/73/i=3/a=036301} {\bibfield
   {journal} {\bibinfo  {journal} {Reports on Progress in Physics}\ }\textbf
  {\bibinfo {volume} {73}},\ \bibinfo {pages} {036301} (\bibinfo {year}
  {2010})}\BibitemShut {NoStop}%
\bibitem [{\citenamefont {Barker}\ and\ \citenamefont
  {Warburton}(1988)}]{Barker1988}%
  \BibitemOpen
  \bibfield  {author} {\bibinfo {author} {\bibfnamefont {F.}~\bibnamefont
  {Barker}}\ and\ \bibinfo {author} {\bibfnamefont {E.}~\bibnamefont
  {Warburton}},\ }\href {\doibase https://doi.org/10.1016/0375-9474(88)90613-6}
  {\bibfield  {journal} {\bibinfo  {journal} {Nuclear Physics A}\ }\textbf
  {\bibinfo {volume} {487}},\ \bibinfo {pages} {269 } (\bibinfo {year}
  {1988})}\BibitemShut {NoStop}%
\bibitem [{\citenamefont {Brune}(2002)}]{Brune2002}%
  \BibitemOpen
  \bibfield  {author} {\bibinfo {author} {\bibfnamefont {C.~R.}\ \bibnamefont
  {Brune}},\ }\href {\doibase 10.1103/PhysRevC.66.044611} {\bibfield  {journal}
  {\bibinfo  {journal} {Phys. Rev. C}\ }\textbf {\bibinfo {volume} {66}},\
  \bibinfo {pages} {044611} (\bibinfo {year} {2002})}\BibitemShut {NoStop}%
\bibitem [{\citenamefont {Wilkinson}\ and\ \citenamefont
  {Macefield}(1974)}]{Wilkinson1974}%
  \BibitemOpen
  \bibfield  {author} {\bibinfo {author} {\bibfnamefont {D.}~\bibnamefont
  {Wilkinson}}\ and\ \bibinfo {author} {\bibfnamefont {B.}~\bibnamefont
  {Macefield}},\ }\href {\doibase https://doi.org/10.1016/0375-9474(74)90645-9}
  {\bibfield  {journal} {\bibinfo  {journal} {Nuclear Physics A}\ }\textbf
  {\bibinfo {volume} {232}},\ \bibinfo {pages} {58 } (\bibinfo {year}
  {1974})}\BibitemShut {NoStop}%
\bibitem [{\citenamefont {Michel}(2007)}]{Michel2007}%
  \BibitemOpen
  \bibfield  {author} {\bibinfo {author} {\bibfnamefont {N.}~\bibnamefont
  {Michel}},\ }\href {\doibase https://doi.org/10.1016/j.cpc.2006.10.004}
  {\bibfield  {journal} {\bibinfo  {journal} {Computer Physics Communications}\
  }\textbf {\bibinfo {volume} {176}},\ \bibinfo {pages} {232 } (\bibinfo {year}
  {2007})}\BibitemShut {NoStop}%
\bibitem [{\citenamefont {Brun}\ and\ \citenamefont
  {Rademakers}(1997)}]{Brun1997}%
  \BibitemOpen
  \bibfield  {author} {\bibinfo {author} {\bibfnamefont {R.}~\bibnamefont
  {Brun}}\ and\ \bibinfo {author} {\bibfnamefont {F.}~\bibnamefont
  {Rademakers}},\ }\href {\doibase 10.1016/S0168-9002(97)00048-X} {\bibfield
  {journal} {\bibinfo  {journal} {Nuclear Instruments and Methods in Physics
  Research Section A: Accelerators, Spectrometers, Detectors and Associated
  Equipment}\ }\textbf {\bibinfo {volume} {389}},\ \bibinfo {pages} {81 }
  (\bibinfo {year} {1997})}\BibitemShut {NoStop}%
\bibitem [{\citenamefont {Baker}\ and\ \citenamefont
  {Cousins}(1984)}]{Baker1984}%
  \BibitemOpen
  \bibfield  {author} {\bibinfo {author} {\bibfnamefont {S.}~\bibnamefont
  {Baker}}\ and\ \bibinfo {author} {\bibfnamefont {R.~D.}\ \bibnamefont
  {Cousins}},\ }\href {\doibase https://doi.org/10.1016/0167-5087(84)90016-4}
  {\bibfield  {journal} {\bibinfo  {journal} {Nuclear Instruments and Methods
  in Physics Research}\ }\textbf {\bibinfo {volume} {221}},\ \bibinfo {pages}
  {437 } (\bibinfo {year} {1984})}\BibitemShut {NoStop}%
\bibitem [{\citenamefont {James}(1994)}]{James1994}%
  \BibitemOpen
  \bibfield  {author} {\bibinfo {author} {\bibfnamefont {F.}~\bibnamefont
  {James}},\ }\href@noop {} {\emph {\bibinfo {title} {{MINUIT Function
  Minimization and Error Analysis: Reference Manual Version 94.1}}}}\ (\bibinfo
  {year} {1994})\BibitemShut {NoStop}%
%%CITATION = CERN-D-506;%%
\bibitem [{\citenamefont {Warburton}(1986)}]{Warburton1986}%
  \BibitemOpen
  \bibfield  {author} {\bibinfo {author} {\bibfnamefont {E.~K.}\ \bibnamefont
  {Warburton}},\ }\href {\doibase 10.1103/PhysRevC.33.303} {\bibfield
  {journal} {\bibinfo  {journal} {Phys. Rev. C}\ }\textbf {\bibinfo {volume}
  {33}},\ \bibinfo {pages} {303} (\bibinfo {year} {1986})}\BibitemShut
  {NoStop}%
\bibitem [{\citenamefont {Hardy}\ and\ \citenamefont
  {Towner}(2005)}]{Hardy2005}%
  \BibitemOpen
  \bibfield  {author} {\bibinfo {author} {\bibfnamefont {J.~C.}\ \bibnamefont
  {Hardy}}\ and\ \bibinfo {author} {\bibfnamefont {I.~S.}\ \bibnamefont
  {Towner}},\ }\href {\doibase 10.1103/PhysRevC.71.055501} {\bibfield
  {journal} {\bibinfo  {journal} {Phys. Rev. C}\ }\textbf {\bibinfo {volume}
  {71}},\ \bibinfo {pages} {055501} (\bibinfo {year} {2005})}\BibitemShut
  {NoStop}%
\bibitem [{\citenamefont {Tanabashi}\ \emph {et~al.}(2018)\citenamefont
  {Tanabashi} \emph {et~al.}}]{Tanabashi2018}%
  \BibitemOpen
  \bibfield  {author} {\bibinfo {author} {\bibfnamefont {M.}~\bibnamefont
  {Tanabashi}} \emph {et~al.} (\bibinfo {collaboration} {Particle Data
  Group}),\ }\href {\doibase 10.1103/PhysRevD.98.030001} {\bibfield  {journal}
  {\bibinfo  {journal} {Phys. Rev. D}\ }\textbf {\bibinfo {volume} {98}},\
  \bibinfo {pages} {030001} (\bibinfo {year} {2018})}\BibitemShut {NoStop}%
\bibitem [{\citenamefont {Riisager}(2014)}]{Riisager2014}%
  \BibitemOpen
  \bibfield  {author} {\bibinfo {author} {\bibfnamefont {K.}~\bibnamefont
  {Riisager}},\ }\href {\doibase
  https://doi.org/10.1016/j.nuclphysa.2014.02.003} {\bibfield  {journal}
  {\bibinfo  {journal} {Nuclear Physics A}\ }\textbf {\bibinfo {volume}
  {925}},\ \bibinfo {pages} {112 } (\bibinfo {year} {2014})}\BibitemShut
  {NoStop}%
\bibitem [{\citenamefont {Teichmann}\ and\ \citenamefont
  {Wigner}(1952)}]{Teichmann1952}%
  \BibitemOpen
  \bibfield  {author} {\bibinfo {author} {\bibfnamefont {T.}~\bibnamefont
  {Teichmann}}\ and\ \bibinfo {author} {\bibfnamefont {E.~P.}\ \bibnamefont
  {Wigner}},\ }\href {\doibase 10.1103/PhysRev.87.123} {\bibfield  {journal}
  {\bibinfo  {journal} {Phys. Rev.}\ }\textbf {\bibinfo {volume} {87}},\
  \bibinfo {pages} {123} (\bibinfo {year} {1952})}\BibitemShut {NoStop}%
\bibitem [{\citenamefont {Borge}\ \emph {et~al.}(2013)\citenamefont {Borge},
  \citenamefont {Fraile}, \citenamefont {Fynbo}, \citenamefont {Jonson},
  \citenamefont {Kirsebom}, \citenamefont {Nilsson}, \citenamefont {Nyman},
  \citenamefont {Possnert}, \citenamefont {Riisager},\ and\ \citenamefont
  {Tengblad}}]{Borge2013}%
  \BibitemOpen
  \bibfield  {author} {\bibinfo {author} {\bibfnamefont {M.~J.~G.}\
  \bibnamefont {Borge}}, \bibinfo {author} {\bibfnamefont {L.~M.}\ \bibnamefont
  {Fraile}}, \bibinfo {author} {\bibfnamefont {H.~O.~U.}\ \bibnamefont
  {Fynbo}}, \bibinfo {author} {\bibfnamefont {B.}~\bibnamefont {Jonson}},
  \bibinfo {author} {\bibfnamefont {O.~S.}\ \bibnamefont {Kirsebom}}, \bibinfo
  {author} {\bibfnamefont {T.}~\bibnamefont {Nilsson}}, \bibinfo {author}
  {\bibfnamefont {G.}~\bibnamefont {Nyman}}, \bibinfo {author} {\bibfnamefont
  {G.}~\bibnamefont {Possnert}}, \bibinfo {author} {\bibfnamefont
  {K.}~\bibnamefont {Riisager}}, \ and\ \bibinfo {author} {\bibfnamefont
  {O.}~\bibnamefont {Tengblad}},\ }\href
  {http://stacks.iop.org/0954-3899/40/i=3/a=035109} {\bibfield  {journal}
  {\bibinfo  {journal} {Journal of Physics G: Nuclear and Particle Physics}\
  }\textbf {\bibinfo {volume} {40}},\ \bibinfo {pages} {035109} (\bibinfo
  {year} {2013})}\BibitemShut {NoStop}%
\bibitem [{\citenamefont {Borge}\ \emph {et~al.}(1997)\citenamefont {Borge}
  \emph {et~al.}}]{Borge1997}%
  \BibitemOpen
  \bibfield  {author} {\bibinfo {author} {\bibfnamefont {M.~J.~G.}\
  \bibnamefont {Borge}} \emph {et~al.},\ }\href {\doibase
  10.1103/PhysRevC.55.R8} {\bibfield  {journal} {\bibinfo  {journal} {Phys.
  Rev. C}\ }\textbf {\bibinfo {volume} {55}},\ \bibinfo {pages} {R8} (\bibinfo
  {year} {1997})}\BibitemShut {NoStop}%
\bibitem [{\citenamefont {Riisager}\ \emph {et~al.}(2015)\citenamefont
  {Riisager}, \citenamefont {Fynbo}, \citenamefont {Hyldegaard},\ and\
  \citenamefont {Jensen}}]{Riisager2015}%
  \BibitemOpen
  \bibfield  {author} {\bibinfo {author} {\bibfnamefont {K.}~\bibnamefont
  {Riisager}}, \bibinfo {author} {\bibfnamefont {H.}~\bibnamefont {Fynbo}},
  \bibinfo {author} {\bibfnamefont {S.}~\bibnamefont {Hyldegaard}}, \ and\
  \bibinfo {author} {\bibfnamefont {A.}~\bibnamefont {Jensen}},\ }\href
  {\doibase https://doi.org/10.1016/j.nuclphysa.2015.04.003} {\bibfield
  {journal} {\bibinfo  {journal} {Nuclear Physics A}\ }\textbf {\bibinfo
  {volume} {940}},\ \bibinfo {pages} {119 } (\bibinfo {year}
  {2015})}\BibitemShut {NoStop}%
\bibitem [{\citenamefont {Sagawa}\ \emph {et~al.}(1993)\citenamefont {Sagawa},
  \citenamefont {Hamamoto},\ and\ \citenamefont {Ishihara}}]{Sagawa1993}%
  \BibitemOpen
  \bibfield  {author} {\bibinfo {author} {\bibfnamefont {H.}~\bibnamefont
  {Sagawa}}, \bibinfo {author} {\bibfnamefont {I.}~\bibnamefont {Hamamoto}}, \
  and\ \bibinfo {author} {\bibfnamefont {M.}~\bibnamefont {Ishihara}},\ }\href
  {\doibase https://doi.org/10.1016/0370-2693(93)91422-J} {\bibfield  {journal}
  {\bibinfo  {journal} {Physics Letters B}\ }\textbf {\bibinfo {volume}
  {303}},\ \bibinfo {pages} {215 } (\bibinfo {year} {1993})}\BibitemShut
  {NoStop}%
\bibitem [{\citenamefont {Kanada-En'yo}\ \emph {et~al.}(1995)\citenamefont
  {Kanada-En'yo}, \citenamefont {Horiuchi},\ and\ \citenamefont
  {Ono}}]{KanadaEnyo1995}%
  \BibitemOpen
  \bibfield  {author} {\bibinfo {author} {\bibfnamefont {Y.}~\bibnamefont
  {Kanada-En'yo}}, \bibinfo {author} {\bibfnamefont {H.}~\bibnamefont
  {Horiuchi}}, \ and\ \bibinfo {author} {\bibfnamefont {A.}~\bibnamefont
  {Ono}},\ }\href {\doibase 10.1103/PhysRevC.52.628} {\bibfield  {journal}
  {\bibinfo  {journal} {Phys. Rev. C}\ }\textbf {\bibinfo {volume} {52}},\
  \bibinfo {pages} {628} (\bibinfo {year} {1995})}\BibitemShut {NoStop}%
\bibitem [{\citenamefont {Teeters}\ and\ \citenamefont
  {Kurath}(1977)}]{Teeters1977}%
  \BibitemOpen
  \bibfield  {author} {\bibinfo {author} {\bibfnamefont {W.}~\bibnamefont
  {Teeters}}\ and\ \bibinfo {author} {\bibfnamefont {D.}~\bibnamefont
  {Kurath}},\ }\href {\doibase https://doi.org/10.1016/0375-9474(77)90275-5}
  {\bibfield  {journal} {\bibinfo  {journal} {Nuclear Physics A}\ }\textbf
  {\bibinfo {volume} {275}},\ \bibinfo {pages} {61 } (\bibinfo {year}
  {1977})}\BibitemShut {NoStop}%
\bibitem [{\citenamefont {Cohen}\ and\ \citenamefont
  {Kurath}(1965)}]{Cohen1965}%
  \BibitemOpen
  \bibfield  {author} {\bibinfo {author} {\bibfnamefont {S.}~\bibnamefont
  {Cohen}}\ and\ \bibinfo {author} {\bibfnamefont {D.}~\bibnamefont {Kurath}},\
  }\href {\doibase https://doi.org/10.1016/0029-5582(65)90148-3} {\bibfield
  {journal} {\bibinfo  {journal} {Nuclear Physics}\ }\textbf {\bibinfo {volume}
  {73}},\ \bibinfo {pages} {1 } (\bibinfo {year} {1965})}\BibitemShut {NoStop}%
\end{thebibliography}%
\end{document}